\documentclass[prb,aps,notitlepage,nofootinbib,superscriptaddress,twocolumn,longbibliography,10pt]{revtex4-2}
\usepackage{amssymb, amsmath}
\usepackage{makecell}
\usepackage[colorlinks=true,citecolor=blue,linkcolor=blue]{hyperref}
\usepackage{graphicx}

\newcommand{\be}{\begin{equation}}
\newcommand{\ee}{\end{equation}}
\newcommand{\ef}{E_\text{F}}
\newcommand{\kf}{k_\text{F}}
\newcommand{\beql}{B_\text{EQL}}
\newcommand{\sah}{\sigma^\text{AH}}

\begin{document}
\begin{titlepage}

\title{Design Principles for Topological Thermoelectrics}

\author{Brian Skinner}
\affiliation{Department of Physics, The Ohio State University, Columbus, 43210, USA}

\author{Poulomi Chakraborty}
\affiliation{Department of Physics, The Ohio State University, Columbus, 43210, USA}

\author{Joshua Scales}
\affiliation{Department of Physics, The Ohio State University, Columbus, 43210, USA}

\author{Joseph P. Heremans}
\affiliation{Department of Physics, The Ohio State University, Columbus, 43210, USA}
\affiliation{Department of Materials Science and Engineering, The Ohio State University, Columbus, 43210, USA }
\affiliation{Department of Mechanical and Aerospace Engineering, The Ohio State University, Columbus, 43210, USA }

\makeatletter

\date{\today}

\begin{abstract}
Conventional metals, insulators, and semimetals are constrained by fundamental limitations in terms of their thermoelectric performance. Topological materials offer certain features that allow them to circumvent these constraints, and potentially to form the basis for thermoelectric devices with unprecedented efficiency. In this article we review the thermoelectric performance of topological materials, focusing specifically on nodal semimetals, such as Weyl and nodal-line semimetals. We discuss how certain unique ``topological'' features of these materials -- namely their topologically protected band touching points, electron-hole degenerate lowest Landau level, and Berry curvature -- allow them to exhibit thermoelectric properties that go beyond what is possible in conventional materials, particularly in the presence of an applied magnetic field. We focus our discussion on the goal of achieving large figure of merit $zT$, and for each material class we summarize optimal \emph{design principles} for selecting materials that maximize thermoelectric efficiency. 

We then use these optimal design principles to design and implement a high-throughput database search for topological semimetals that are promising as thermoelectrics. In addition to highlighting a number of materials that are already known to have large magnetothermoelectric effects, our search uncovers twelve additional materials that are especially promising for near-future experiments.
\end{abstract}

\maketitle
\vspace{2mm}
\end{titlepage}

\section{Introduction}

The thermoelectric effect is the direct conversion of a temperature difference $\Delta T$ across a material to a voltage difference $\Delta V$. It arises because the diffusion constant of mobile electrons or holes within the material is generally an increasing function of temperature, so electrons or holes on the hot side of a material rapidly diffuse away and accumulate on the cold side, where they are slow to diffuse back. The accumulation of carriers on the cold side sets up a voltage difference whose sign reflects the sign of the carriers.

This seemingly mundane effect has potentially enormous practical implications, since thermoelectricity represents a direct conversion between heat and electric power without moving parts. Heat engines that accomplish similar conversion generally involve a working fluid that undergoes mechanical compression and expansion. In thermoelectric processes, the working fluid is the electron system itself. So the thermoelectric effect offers the promise of simple, solid-state heat pumps that provide active heating and cooling or that allow for the recovery of waste heat. 

Unfortunately, the efficiency of thermoelectric processes is typically low, as we discuss in the following section. The characteristic scale of $\Delta V/\Delta T$ is $k_B/e \approx 86$\,$\mu$V/K, where $k_B$ is the Boltzmann constant and $-e$ is the electron charge. So even one hundred Kelvin of temperature difference across a material can generally be expected to produce a voltage no greater than $10$ mV.

Nonetheless, thermoelectric technologies have a number of important use cases \cite{freer_realising_2020}, and research into the thermoelectric effect remains an active field making steady advances, spurred by ongoing advancements in solid state physics and materials science. In this article we consider the relevance of topological materials, which fall outside the traditional dichotomy of all solid materials as either metals or insulators, for thermoelectric applications. We overview the unique benefits provided by topological band structure for both longitudinal and transverse thermoelectric response, and we attempt to summarize optimal \emph{design principles} for maximizing these benefits. We focus particularly on the topological semimetals, for which conduction and valence bands meet at a point or line in momentum space that can serve as a source of Berry curvature \cite{Burkov_Hook_Balents_2011, armitage_weyl_2018}.

Throughout this article, our focus is on providing simple and intuitive arguments that elucidate the mechanisms that drive thermoelectric response in topological semimetals and how they can be optimized, rather than providing precise derivations or a thorough literature overview. For a larger overview of recent progress in thermoelectric materials and devices we refer the reader to the reviews in Refs.~\cite{freer_realising_2020, shakouri_recent_2011, zhang_thermoelectric_2015, fu_topological_2020, yan_high-performance_2022,  sun_strategies_2022, adachi_fundamentals_2025}.

\section{Fundamentals and limitations of thermoelectric response in conventional materials}
\label{sec:conventional} 

Before discussing the unique features enabled by topological band structure, we first summarize the basic properties of thermoelectric response and we overview its fundamental limitations in conventional metals and insulators.

\subsection{Thermopower and Figure of Merit}

In general, a material's thermoelectric response is quantified by the thermoelectric tensor
\be 
S_{ij} = \frac{(\Delta V)_j}{(\Delta T)_i},
\ee
where $i,j \in \{x, y, z\}$ label coordinate directions. When the gradient in voltage is along the same direction $i$ as the gradient in temperature, the coefficient $S_{ii}$ is called the Seebeck coefficient (or simply the ``thermopower'') in the $i$ direction. The sign of $S_{ii}$ generally reflects the sign of the mobile carriers: when the carriers are electrons, their increased concentration on the cold side produces a voltage gradient that is opposite to the temperature gradient, so that $S_{ii}$ is negative; on the other hand, $S_{ii}$ is positive when the carriers are holes. When $j$ is orthogonal to $i$, the quantity $S_{ij}$ is called the Nernst coefficient. The thermoelectric coefficients $S_{ii}$ and $S_{ij}$ are well-defined only in the limit where $\Delta T$ and $\Delta V$ are small enough that $\Delta V \propto \Delta T$, i.e., in the limit of linear response.

While large $S_{ii}$ or $S_{ij}$ is necessary for efficient thermoelectric conversion, it is not sufficient. In order to calculate the efficiency of thermoelectric energy conversion, defined (say, for an electric generator) as the ratio of the electric power generated to the heat flux through the device, one must take into account Joule heating losses inside the material. This calculation (see, e.g., Ref.~\cite{heikes_thermoelectricity:_1961}) shows that, after optimizing the load resistance of the circuit, the thermoelectric efficiency is equal to the Carnot efficiency $(T_h - T_c)/T_h$ (where $T_h$ and $T_c$ are the usual hot-bath and cold-bath temperatures) multiplied by a dimensionless function that depends only on the dimensionless ratio $zT$, defined as
\be 
zT = \frac{S^2 T}{\kappa \rho},
\label{eq:zT}
\ee 
where $\kappa$ is the thermal conductivity (which should be low so that a given heat flux is associated with a large temperature difference) and $\rho$ is the electrical resistance (which should be low in order to minimize Joule heating losses).  The thermoelectric efficiency approaches the Carnot limit as $zT \rightarrow \infty$. For applications using the Seebeck response in the $i$ direction, the value of $zT$ depends only on the diagonal component of each tensor in the $i$ direction ($zT = S_{ii}^2 T /\kappa_{ii} \rho_{ii}$). For applications using the Nernst response, the value of $zT$ becomes
\be 
\left(zT\right)_\textrm{transverse} = \frac{S_{ij}^2 T}{\kappa_{ii} \rho_{jj}},
\label{eq:zTtrans}
\ee 
where $i$ is the direction of the temperature gradient and $j$ is the direction of the voltage gradient. 

It is conceptually useful to think about $S_{ij}$ using the Onsager reciprocal relation between the thermoelectric effect and the Peltier effect, which describes the generation of a heat current by an electric current in conditions of uniform temperature $T$. Specifically,
\be 
S_{ij} = \frac{1}{T} \frac{J^Q_i}{J^e_j},
\label{eq:SdefOnsager}
\ee
where $J^Q_j$ is the heat current density in direction $j$ and $J^e_i$ is the electric current density in direction $i$. In this language, good thermoelectric materials are those for which a small electric current $J^e$ generates a large heat current $J^Q$.

\begin{figure}[htbp]
  \centering
  \includegraphics[width=3.1in]{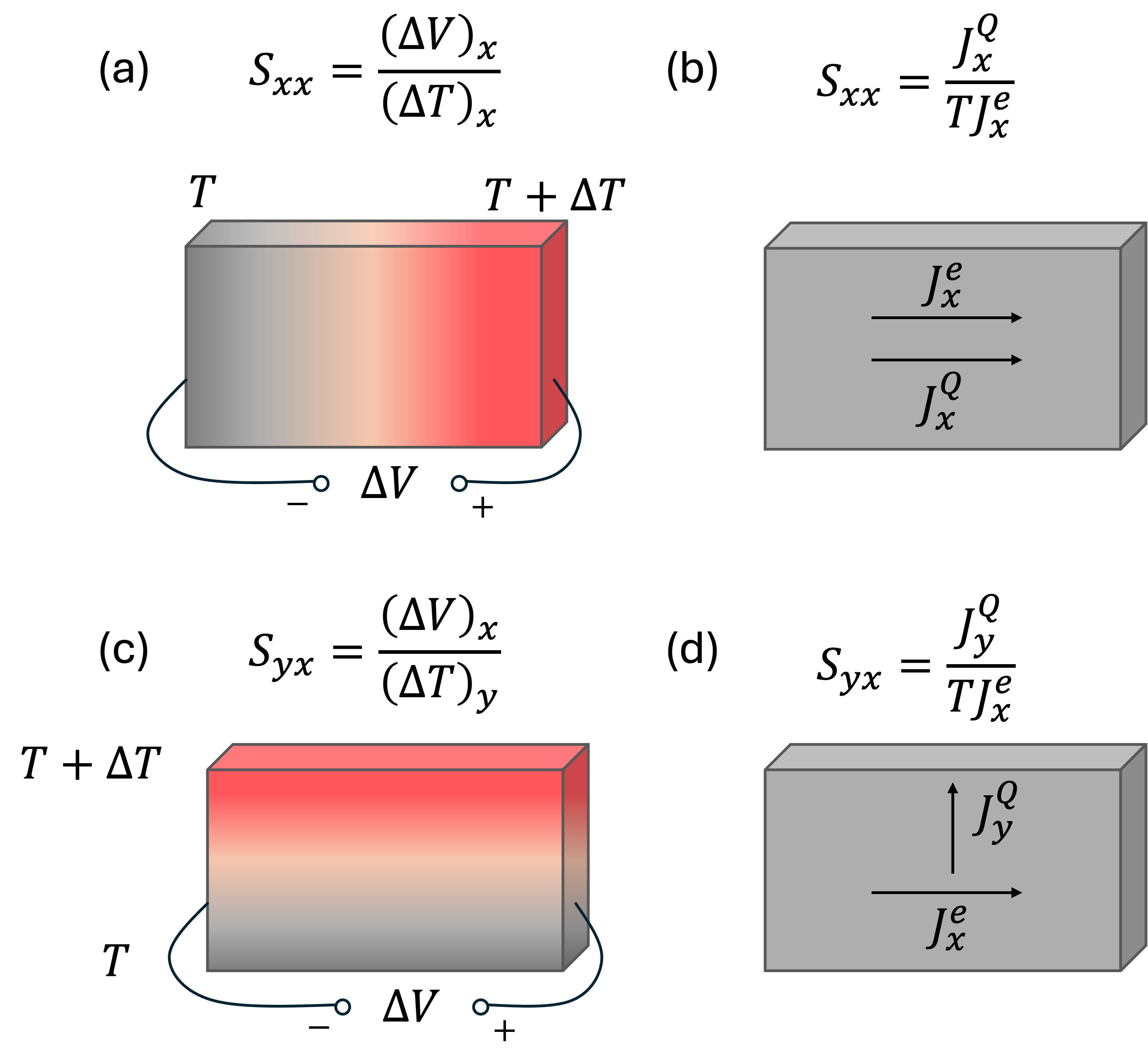}
  \caption{The Seebeck coefficient $S_{xx}$ can be defined either (a) in terms of the ratio $(\Delta V)_x / (\Delta T)_x$ under conditions of zero current, or (b) in terms of the ratio between the heat current and the electric current under conditions of uniform temperature. The Nernst coefficient similarly has two equivalent definitions, depicted in (c) and (d).}
  \label{fig:S-schematics}
\end{figure}

\subsection{Semiclassical calculation procedure}

We now briefly recapitulate the mathematical recipe for calculating the thermoelectric tensor $\hat{S}$ using semiclassical transport equations \cite{Ashcroft_Mermin_1976}. The general equations that govern the electrical and heat currents are
\begin{align}
    \vec{J}^e &= \hat{\sigma}\vec{E} - \hat{\alpha}\vec{\nabla}{T},
    \label{eq:je}
    \\
    \vec{J}^Q &= T\hat{\alpha}\vec{E} - \hat{\kappa}\vec{\nabla}{T},
    \label{eq:jq}
\end{align}
where $\vec{E}$ is the electric field and $\hat{\sigma}$, $\hat{\alpha}$, and $\hat{\kappa}$ are the electrical, Peltier (or ``thermoelectric''), and thermal conductivity tensors. The appearance of the same tensor $\hat{\alpha}$ in both Eqs.~(\ref{eq:je}) and (\ref{eq:jq}) is a reflection of the Onsager reciprocal relation mentioned above. The tensor $\hat{S}$ is defined by the relation $\vec{E} = \hat{S} \vec{\nabla}T$ under conditions where $\vec{J}^e = 0$, so that
\begin{align}
    \hat{S} = \hat{\sigma}^{-1}\hat{\alpha}.
    \label{eq:generalS}
\end{align}
As an example, for situations where a magnetic field is applied in the $z$ direction, Eq.~(\ref{eq:generalS}) implies a Seebeck coefficient
\be 
S_{xx} = \frac{\alpha_{xx} \sigma_{yy} - \alpha_{yx} \sigma_{xy}}{\sigma_{xx} \sigma_{yy} - \sigma_{xy} \sigma_{yx}} = \alpha_{xx} \rho_{xx} + \alpha_{yx} \rho_{xy},
\label{eq:Seebeckdeff}
\ee 
where $\hat{\rho} = \hat{\sigma}^{-1}$ is the resistivity tensor,
and a Nernst coefficient
\be 
S_{yx} = \frac{\alpha_{yx} \sigma_{xx} - \alpha_{xx} \sigma_{yx}}{\sigma_{xx} \sigma_{yy} - \sigma_{xy} \sigma_{yx}} =  \alpha_{xx} \rho_{yx} + \alpha_{yx} \rho_{yy}.
\label{eq:Nernstdeff}
\ee 


At zero temperature, the electrical conductivity tensor is a function of the Fermi energy, $\hat{\sigma} = \hat{\sigma}(\ef)$, and its value is related to the diffusion tensor $\hat{D}$ by the Einstein relation:
\be 
\hat{\sigma}(E) = e^2 g(E) \hat{D}(E).
\ee
In the absence of time reversal symmetry breaking (either through an external magnetic field or magnetic order within the material itself), the off-diagonal components of $\hat{D}$ are zero, while the diagonal components $\hat{D} = (1/3) v_i^2 \tau$, where $v_i$ is the band velocity (at energy $E$) in the $i$ direction and $\tau = \tau(E)$ is the momentum relaxation time. In the presence of a magnetic field, $\hat{D}$ acquires non-diagonal components due to the Lorentz force. For a magnetic field in the $z$ direction, the value of $\sigma_{zz}$ is unaffected by the field and the values of $\sigma_{xz} = \sigma_{yz} = 0$, but the values of $\sigma_{xx}$, $\sigma_{yy}$, and $\sigma_{xy}$ are altered by the magnetic field according to:
\be 
\sigma_{ii}(E) = \frac{1}{3} e^2 g(E) v_i^2 \tau \frac{1}{1 + (\omega_c(E) \tau)^2}, \hspace{5mm} \textrm{for }i, j = x, y,
\label{eq:sigmaxxE}
\ee 
and 
\be 
\sigma_{xy} = \frac{1}{3} e^2 g(E) v_x v_y \tau \frac{ \omega_c(E) \tau }{1 + (\omega_c(E) \tau)^2},
\label{eq:sigmaxyE}
\ee 
where $\omega_c(E)$ is the cyclotron frequency for electrons at energy $E$. For electrons with band mass $m$, the cyclotron frequency is a constant $\omega_c = e B/m$. On the other hand, for electrons having a linear dispersion with velocities $v_x$ and $v_y$ in the $x$ and $y$ directions, respectively, the cyclotron frequency is given by
\be 
\omega_c(E) = \frac{e B v_x v_y}{E}, \hspace{5mm} (\textrm{for Weyl/Dirac electrons}).
\label{eq: omegacE}
\ee 
In general, the off-diagonal elements of the conductivity tensor can be related to each other by the Onsager relation $\sigma_{ij}(B) = \sigma_{ji}(-B)$. (This relation holds more generally any time $B$ is taken to be the direction of time reversal symmetry breaking, so that systems with an anomalous Hall effect due to magnetic order obey a similar relation if $B$ is replaced by the magnetization.) 


The finite temperature conductivity $\hat{\sigma}$ is generally related to the zero-temperature conductivity $\hat{\sigma}(E)$ by a weighted integral over the quasiparticle energy $E$. Namely:
\begin{align}
    \sigma_{ij} &= \int dE \left( -\frac{\partial f}{\partial E} \right) \sigma_{ij} (E) 
    \label{eq: sigma conductivity},
\end{align}
where 
\be
f(E) = \frac{1}{1 + \exp\left( \frac{E - \mu}{k_B T}\right) }
\ee
is the Fermi-Dirac distribution and $\mu$ is the chemical potential.

The Peltier conductivity $\hat{\alpha}$ describes the energy current under the influence of an electric field, and so it is given by a similar integral that is additionally weighted by the energy $(E-\mu)$ of quasiparticles relative to the chemical potential:
\begin{align}
\alpha_{ij} &= \frac{1}{eT} \int dE  (E - \mu) \left( -\frac{\partial f}{\partial E} \right) \sigma_{ij} (E) 
    \label{eq: alpha conductivity}.
\end{align}
In the limit of low temperature $k_B T \ll \mu$, taking a Sommerfeld expansion of the integrals in Eqs.~(\ref{eq: sigma conductivity}) and (\ref{eq: alpha conductivity}) and plugging them into Eq.~(\ref{eq:generalS}) gives the well-known Mott formula:
\begin{align}
    \hat{S} = \frac{k_B}{e}\frac{\pi^2}{3} k_B T \hat{\sigma}(\mu)^{-1} \left.\frac{d\hat{\sigma}(E)}{d E} \right|_{E = \mu}.
    \label{eq:Mott_formula}
\end{align}
Conceptually, one can understand the proportionality between the thermopower and the derivative $d\sigma/dE$ as follows. At finite temperature, the electron system has thermal energy in the form of excited electrons at energies $\sim k_B T$ above $\mu$ and holes at energies $\sim k_BT$ below $\mu$. Under the influence of an electric field, electrons and holes carry their heat in opposite directions. In order for the electron system to carry a significant heat current $J^Q$, the electrons must conduct better than the holes, in the sense of having a higher drift velocity. So the thermopower reflects the difference in their effective conductivity $\sigma_\textrm{electrons} - \sigma_\textrm{holes} \sim (k_B T) d\sigma/dE$.

\subsection{Limitations on thermoelectric performance of conventional materials}

In a conductor for which the thermal energy $k_B T$ is much lower than the Fermi energy $\ef$, one can quickly understand the magnitude of the Seebeck coefficient using the following semiquantitative argument. (Here and below, we use the term ``Fermi energy'' to refer to the chemical potential at zero temperature, and we use the more general term ``chemical potential'' to discuss the temperature-dependent quantity.) Under the influence of an electric field $E_x$ in the $x$ direction, the electric current density $J^e_x = n e v_x$, where $n$ is the electron concentration and $v_x$ is the electric-field-induced drift velocity. The net drift of the electron system also implies a heat current that is proportional to the thermal energy density $\sim (k_B T)^2 g(\ef)$, where $g(\ef)$ is the density of states at the Fermi energy. ($k_BT$ is the energy per electron-hole excitation near the Fermi level, and $k_BT g(\ef)$ represents the number of such excitations per unit volume.) Thus the heat current $J^Q_x \sim (k_BT)^2 g(\ef) v_x$, and using Eq.~(\ref{eq:SdefOnsager}) we arrive at $S_{xx} \sim (k_B/e) k_B T g(\ef)/n$. For a typical electron band, $g(\ef)$ is of the order of $n/E_F$, so that $S_{xx} \sim (k_B/e)(k_B T/\ef)$. For a stoichiometric metal, $\ef/k_B$ is typically on the order of $10^5$\ K, so that such metals generically have very small thermopower. Put more succinctly, in a good metal the thermopower is small because all electrons in the band can carry current, while only those within $\sim k_B T$ of the Fermi energy can carry heat. For this basic reason thermoelectrics are almost universally made from semiconductors, for which the value of $\ef$ can be controlled and made small by selective doping.

As the Fermi energy is reduced (say, by reducing the doping), the thermopower $S$ at a given temperature increases until $\ef$ is low enough that $k_B T \gtrsim \ef$. At such low Fermi energy (relative to the temperature), one can think that the electron system comprises something like a classical thermal gas, in which all carriers have an energy of order $k_B T$. Consequently the heat current under the influence of an electric field is of order $n k_B T v_x$, and $S = J^Q/(T J^e) \sim k_B /e$. If the doping is made very low, on the other hand, a doped semiconductor undergoes a metal-to-insulator transition and the chemical potential sinks into the band gap, such that free, delocalized electrons are exponentially rare. This insulating state can produce large $S$ (since rare free carriers have lots of thermal energy, having been thermally activated from the chemical potential to the band edge), but this large thermopower comes at the cost of an exponentially large resistivity $\rho$, so that $zT$ is also exponentially small. Thus, for traditional semiconductors, the optimal doping for thermoelectric efficiency occurs when $k_B T$ and $\mu$ (relative to the band edge) are of the same order of magnitude \cite{mahan_figure_1989, sofo_optimum_1994}.

At this optimal doping, the figure of merit $zT \sim (k_B/e)^2 / L$, where $L \equiv \kappa \rho / T$ is a dimensionless ratio. In the idealized case where there is no contribution to the thermal conductivity from any source beside the electrons themselves, $L$ is called the Lorenz number and its value is a constant of order $(k_B/e)^2$. (For Fermi liquids, $L = (\pi^2/3)(k_B/e)^2$ is called the Wiedemann-Franz law). Including other sources of thermal conductivity, such as phonons (which are usually dominant over the electron contribution when the Fermi energy is low), only increases $L$ and drives $zT$ to a smaller value. Consequently, one can conclude that at optimal doping and in the absence of phonons (or other heat-carrying excitations), $zT$ is at best an order-1 constant \cite{mahan_figure_1989} -- either smaller or larger doping leads to smaller $zT$. For this fundamental reason, achieving $zT \gg 1$ in semiconductors is difficult, and progress in $zT$ over time has been relatively slow. For example, the highest reported value of $zT$ across all materials and temperatures increased only from $\sim 0.5$ to $\sim 2.8$ from 1960 to 2020 \cite{zhang_thermoelectric_2015, sun_strategies_2022}. 

Various ideas have been proposed over time for circumventing this fundamental limitation and achieving large $zT$. For example, one can look for electron bands that have a sharp feature in the density of states $g(E)$ or in the scattering rate $1/\tau(E)$ \cite{dresselhaus_new_2007}. Arranging, through doping, for the Fermi energy to coincide with this sharp feature may have the effect of producing large heat current for a given electrical current [or, in the language of the Mott Formula, Eq.~(\ref{eq:Mott_formula}), a large derivative $d\sigma/dE$]. Quantum confinement of the electron system into nanostructures, which has the effect of producing quantized energy spectra with sharp peaks in the density of states, is a much-studied approach \cite{dresselhaus_low-dimensional_1999, shakouri_recent_2011}.

On the other hand, materials with topological band structure offer three new ingredients that go beyond the preceding arguments about semiconductors and their fundamental limitations. We summarize these ingredients as:
\begin{enumerate}
    \item[1.] \textbf{\textit{Topologically protected band touching points}}.  A topological semimetal has band touching points in the bulk of the material, while a topological insulator has band touching points only on its surface. These band touching points allow one to access regimes of very small Fermi energy without worrying about the system becoming an insulator. That is, a topological material can remain conducting (either in its bulk or on its surface) at any value of the doping.

    \item[2.] \textbf{\textit{An electron-hole degenerate lowest Landau level}}. In general, in the presence of a magnetic field, electron states within both conduction and valence bands have their motion transverse to the magnetic field quantized into discrete energy levels called Landau levels. Topological semimetals have the feature that both conduction and valence bands share a Landau level at an energy precisely equal to the band touching energy. As we show below, this degeneracy enables a potentially huge increase in the thermopower at sufficiently strong magnetic field.

    \item[3.] \textbf{\textit{Quantum geometry, including nontrivial Berry curvature and quantum metric}}. The motion of electrons can exhibit ``anomalous'' responses driven by the nontrivial ``quantum geometric'' structure of the electron wave functions. These responses can mimic the effects of a magnetic field in the sense of providing a component of the electric current that flows transverse to an applied electric field.
\end{enumerate}

Our goal in this article is to give an overview of the way in which these three ingredients can provide enhancement of longitudinal (Sec.\ \ref{sec:longitudinal}) and transverse (Sec.\ \ref{sec:transverse}) thermoelectric effects.

\subsection{A comment on phonon and magnon drag}

In our discussion so far we have assumed that the thermoelectric effect is driven only by the response of electrons and holes. In the language of Eq.~(\ref{eq:SdefOnsager}), we are assuming that when an electric current density $J^e$ is present, the resulting heat current density $J^Q$ is only that of the electron system. In principle, however, \emph{any} excitation in a solid system can carry heat. So if the electron system imparts some of its momentum to, say, the phonon system, then these phonons can also contribute to the heat current density and therefore enhance the thermopower.

This effect is known as ``phonon drag'', and in certain situations it can produce a huge enhancement of the thermopower. A clear expository introduction to phonon drag is given, e.g., in Ref.~\cite{Behnia_2015}, but one can quickly understand its importance in a semiquantitative way by imagining that when the electron system acquires a drift velocity $v_\text{e}$, the process of electron-phonon scattering imparts a drift velocity $v_\text{ph} \propto v_e$ due to electron-phonon scattering. The resulting phonon heat current $J^Q_\text{ph} \sim T C_\text{ph} v_\text{ph}$, where $C_\text{ph}$ is the specific heat of the phonon system. Since the electrical current $J^e = e n v_\text{e}$, we arrive at a phonon-drag contribution to the thermopower
\be 
S^\text{phonon drag} \sim \frac{C_\text{ph}}{e n} \frac{v_\text{ph}}{v_{\text{e}}},
\label{eq:Sphonondrag}
\ee 
where the ratio $v_\text{ph}/v_{\text{e}}$ is a dimensionless number that depends on the rate of electron-phonon scattering and on the rate at which the phonon system loses its momentum to other processes (like phonon-impurity scattering or phonon-phonon umklapp scattering).

In this review we are mostly neglecting the effects of phonon drag, focusing instead on the purely electronic contribution. In real materials it is usually not easy to clearly establish the phonon drag contribution to the observed Seebeck signal. However, it is clear from the simple considerations leading to Eq.~(\ref{eq:Sphonondrag}) that for materials with a sufficiently large phonon specific heat and sufficiently low electronic carrier concentration the effect of phonon drag can be significant. In magnetic materials, there is an analogous effect of magnon drag, which has been claimed to dominate the Seebeck response in the elemental ferromagnets Fe, Ni, and Co \cite{blatt_magnon-drag_1967, watzman_magnon-drag_2016}.

\section{Longitudinal (Seebeck) Effects}
\label{sec:longitudinal}

In this section we overview the behavior of the Seebeck coefficient and the corresponding figure of merit $zT$ for topological materials. For each material type we summarize optimal design criteria for achieving large $zT$, and we conclude the section with a summary of experimental measurements of the Seebeck effect in topological semimetals.

\subsection{Dirac and Weyl semimetals at zero magnetic field}
\label{sec:WeylSxxB0}

\begin{figure}[htbp]
  \centering
  \includegraphics[width=3.1in]{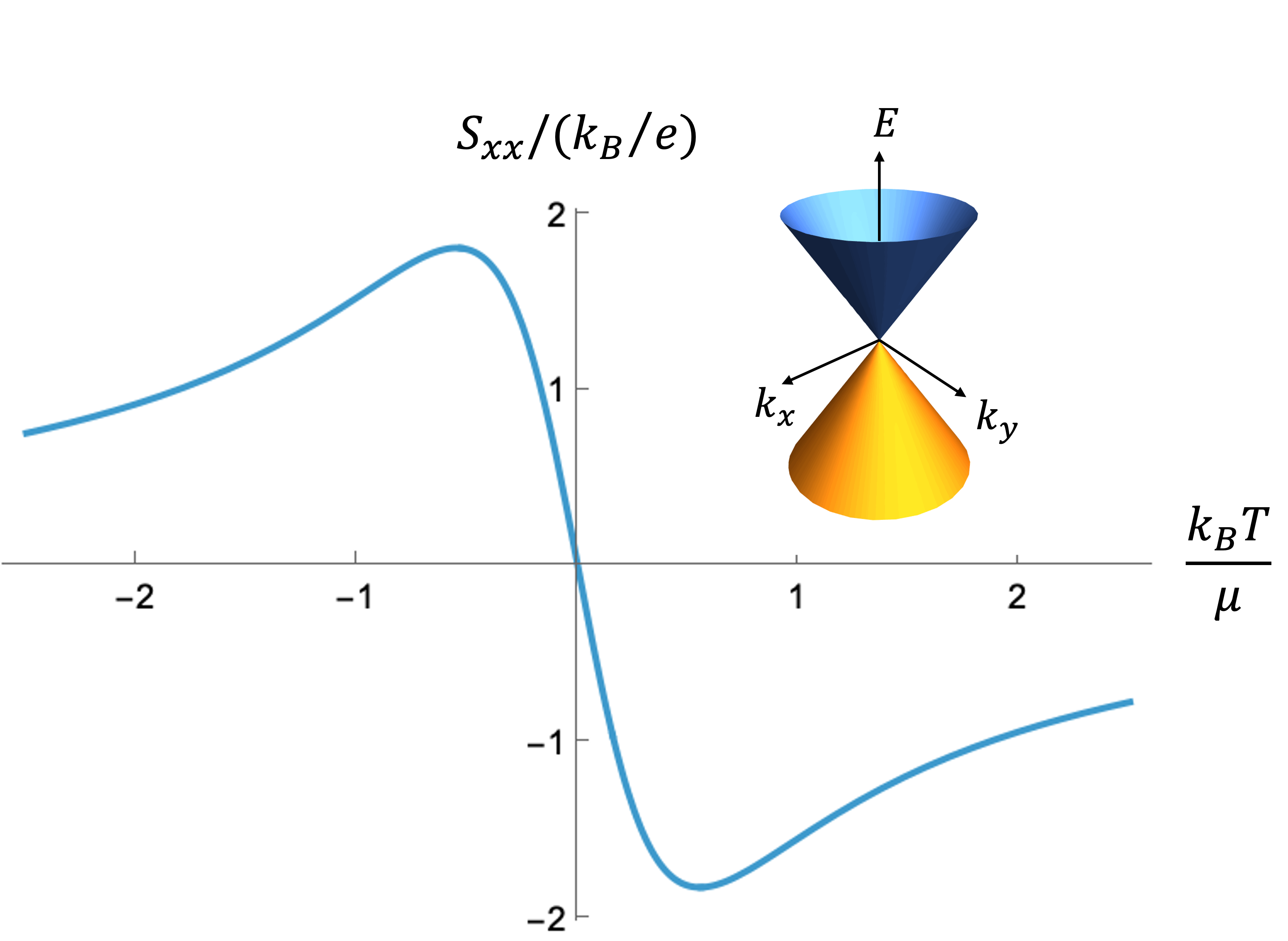}
  \caption{The Seebeck coefficient $S_{xx}$ of a non-magnetic Weyl or Dirac semimetal in the absence of magnetic field is plotted as a function of $k_B T / \mu$, where $\mu$ is the chemical potential [see Eq.~(\ref{eq:Sweylsimple})]. Details of the scattering mechanism can change the precise values of this curve (and $\mu$ in general is temperature-dependent at higher temperatures), but the appearance of a peak $S_{xx} \sim k_B/e$ at $k_B T \sim \mu$ is universal. The inset shows the Dirac/Weyl dispersion relation, with the conduction band (blue) and valence band (orange) meeting at a point in momentum space.}
  \label{fig:Weyl-dispersion}
\end{figure}

A Weyl semimetal is a topological semimetal for which the conduction and valence bands meet at a point in momentum space and form a cone. The low-energy dispersion relation
\be 
E_{\pm}(\vec{k}) = \pm \hbar \left| \vec{v} \cdot \vec{k} \right|,
\ee 
where $\hbar$ is the reduced Planck constant, $\vec{v} = (v_x, v_y, v_z)$ is the set of Weyl velocities, $\vec{k}$ is the wave vector relative to the band touching point, and $E$ is the energy relative to the band touching point. 
In this review we focus our attention primarily on the so-called type-I Weyl semimetals, for which all three velocities $v_x, v_y, v_z$ are positive (depicted in the inset of Fig.~\ref{fig:Weyl-dispersion}). Having a negative velocity in some direction implies that both the conduction and valence band pass through $E=0$, generating potentially large pockets of electron and hole carriers and leading to large typical values of $\ef$. Such large Fermi energy tends to reduce the thermopower, as discussed above. We comment further on such type-II semimetals in Sec.~\ref{sec:compensated}.

The density of states of a (type-I) Weyl semimetal increases parabolically with energy,
\be 
g(E) = \frac{g_0 E^2}{2 \pi^2 \hbar^3 v_x v_y v_z}.
\ee 
The quantity $g_0$ is an integer that represents the degeneracy (spin $\times$ valley) of the Weyl point. The Fermi energy is related to the concentration of electrons $n$ via $\ef = \hbar (v_x v_y v_z)^{1/3} (6 \pi^2 n/g_0)^{1/3}$. Our discussion in this section does not make any reference to any features of quantum geometry (e.g., the Berry curvature), so that our results describe both Weyl semimetals (which are topologically nontrivial) and gapless Dirac semimetals (which are topologically trivial); for the purposes of this discussion, Dirac and Weyl semimetals are different only in that Dirac semimetals have a twice-larger value of $g_0$ per band touching point. We also focus throughout this paper only on bulk responses, so that the thermoelectric contribution arising from Fermi arc states in Weyl semimetals \cite{mccormick_fermi_2018} is not relevant. Quantum geometric effects (in the form of an anomalous Hall conductivity) can play a significant role in the Nernst effect, as we discuss in Sec.~\ref{sec:transverse}.

In the absence of magnetic field, the Seebeck coefficient $S_{xx}$ of a Weyl semimetal can be evaluated using Eqs.~(\ref{eq:generalS})--(\ref{eq: alpha conductivity}), together with a diffusion constant $D_{xx} = (1/3) v_x^2 \tau$. If we assume, as a simple model, that the transport scattering time $\tau$ is an energy-independent constant, then this calculation gives \cite{scott_doping_2023}
\be 
S_{xx} = - \frac{k_B}{e} \frac{2 \pi^2 k_B T \mu}{3 \mu^2 + \pi^2(k_B T)^2 }.
\label{eq:Sweylsimple}
\ee 
This dependence is shown in Fig.~\ref{fig:Weyl-dispersion}: as a function of increasing temperature $k_BT/\mu$, the Seebeck coefficient first rises linearly in $T$ before collapsing again when $k_B T$ is large enough to create a significant population of holes in the valence band. These holes cancel the contribution of conduction band electrons in the heat current, bringing the value of $S_{xx}$ back toward zero. The corresponding peak of $S_{xx}$ occurs when $k_BT$ is of the same order as $\mu$ and the peak value of $S_{xx}$ is of order $k_B /e$. Introducing a dependence of $\tau$ on the quasiparticle energy $E$ changes various numerical prefactors in Eq.~(\ref{eq:Sweylsimple}), but does not change these two general features of the curve $S_{xx}(T)$. An additional complication is that the chemical potential $\mu$ is generally temperature-dependent and tends to sink toward points of minimal density of states (e.g., the band touching point) as the temperature is increased. However, the general features shown in Fig.~\ref{fig:Weyl-dispersion} are unchanged so long as the conductivity is dominated by carriers in the Weyl band (rather than some other trivial band). Notice that there is no explicit dependence of the peak value of $S_{xx}$ on the band parameters (e.g., on the Weyl velocity) or on the scattering time (the electron mobility).

When the temperature and Fermi energy are aligned so that $S_{xx}$ is near its peak value, the electrical resistance $\rho_{xx} \sim 1/[e^2 v_x^2 \tau g(\mu)]$, so that
\be 
(zT)_\text{optimal} \sim \frac{k_B^2 T v_x^2 \tau g(\mu)}{\kappa_{xx}},
\label{eq:zTWeylB0}
\ee 
with $\mu \sim k_BT$. This expression is simplified considerably if one recalls that the scattering rate $1/\tau(E)$ is generally proportional to the electronic density of states $g(E)$ (as dictated by Fermi's golden rule), so that $\tau g$ is a constant (which may be temperature-dependent due to phonon scattering, but does not depend on band parameters). We then arrive at a simple result
\be 
(zT)_\text{optimal} \propto \frac{v_x^2}{\kappa_{xx}}.
\ee 

In terms of optimal design of thermoelectrics at a fixed temperature, one can summarize the results of this subsection as follows:

\

\noindent \fbox{%
\parbox{0.96\linewidth}{ \textbf{Optimal design criteria for longitudinal thermoelectric response of Dirac and Weyl semimetals at $B=0$:}
\begin{itemize}
    \item The doping should be such that the Fermi energy is comparable to $k_B T$.
    \item The electron velocity $v_x$ in the transport direction should be high and the lattice thermal conductivity $\kappa_{xx}$ should be low so as to maximize the quantity $v_x^2/\kappa_{xx}$.
    \item The material should be as little disordered as possible, which reduces $\rho_{xx}$. 
    \item There should be no (or minimal) trivial bands coexisting in energy with the Weyl bands, so that the conductivity (in the transport direction) is dominated by the Weyl bands.
\end{itemize}
} }

\

As mentioned above, however, even when these criteria are satisfied the value of $(zT)_\textrm{optimal}$ cannot be parametrically larger than unity, since the electronic contribution to the thermal conductivity is always large enough that $\kappa \rho / T \gtrsim (k_B/e)^2$.

\subsection{Dirac and Weyl semimetals in a magnetic field}
\label{sec:WeylSxxB}

For an electron system in a magnetic field, there are two intrinsic scales of magnetic field. First, there is a field scale $B_1$ corresponding to $\omega_c \tau = 1$. For electrons with band mass $m$, the value of $B_1$ is given simply by $B_1 = m/(e \tau)$. In Weyl semimetals, however, the cyclotron frequency $\omega_c$ is energy-dependent [see Eq.~(\ref{eq: omegacE})], so that $B_1$ also depends on the electron energy, 
\be 
B_1(E) = \frac{E}{e \tau v_x v_y}
\ee 
(for a magnetic field applied in the $z$ direction). 
In this section, when we refer to $B_1$ we mean its value at $E = \ef$, and we focus on the regime of sufficiently low temperature that $k_B T \lesssim \ef$. 
In terms of the semiclassical equations for the conductivity [Eqs.~(\ref{eq:sigmaxxE}), (\ref{eq:sigmaxyE})], $B_1$ has the meaning that $\sigma_{xx} \gg \sigma_{xy}$ (and similarly $\rho_{xx} \ll \rho_{xy}$) when $B \ll B_1$, or in other words at $B \ll B_1$ electrons mostly drift in the direction of the (negative) electric field. At $B \gg B_1$, on the other hand, electrons mostly drift perpendicular to the electric field. 

\begin{figure}[htbp]
  \centering
  \includegraphics[width=3.1in]{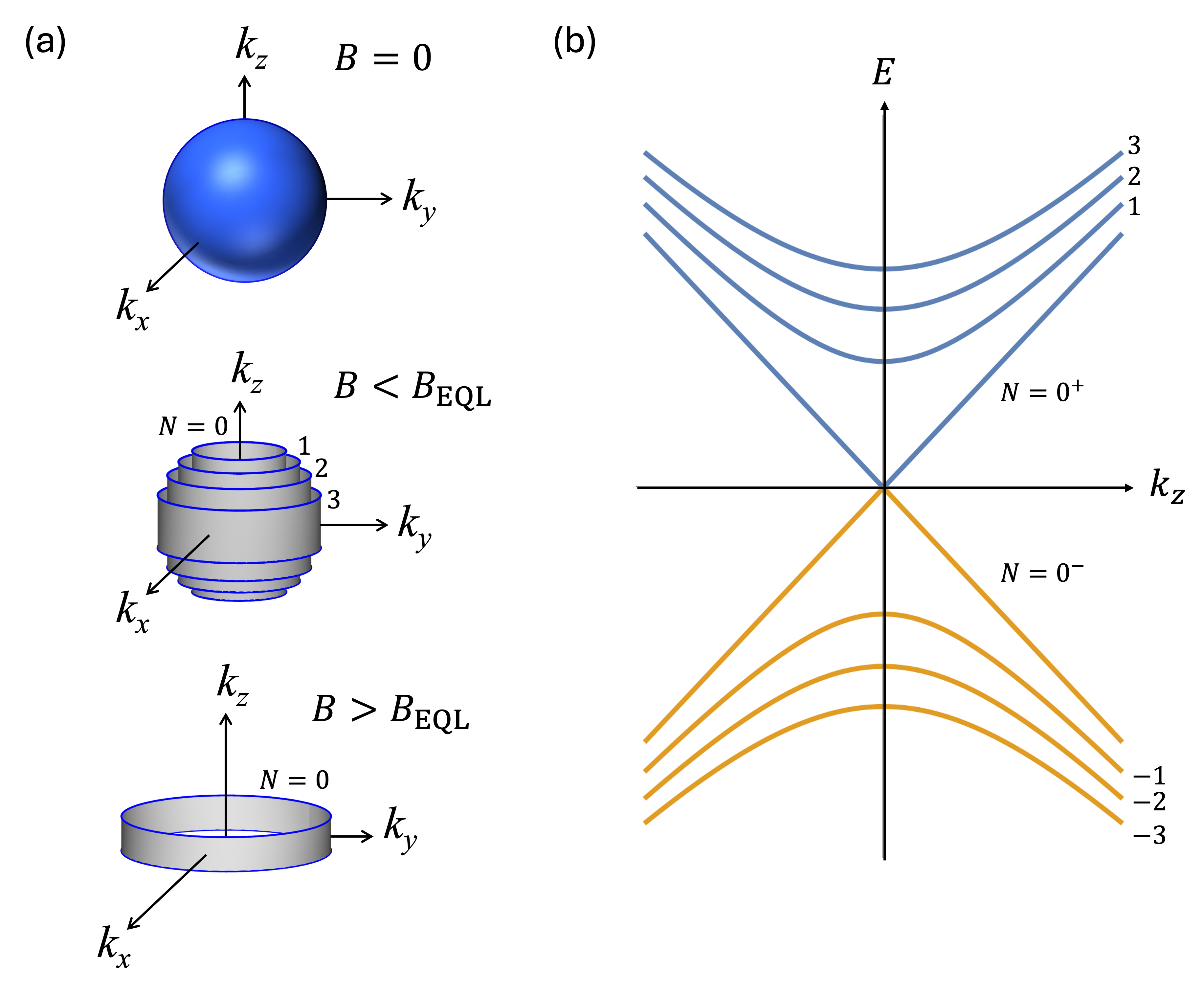}
  \caption{(a) The evolution of the Fermi surface is depicted as a function of magnetic field $B$ in the $z$ direction. A spherical Fermi surface is broken up into discrete ``cylinders'' associated with constant kinetic energy in the transverse direction. In the EQL, only one such cylinder is occupied, and the dispersion becomes effectively one-dimensional. (b) The dispersion relation $E(k_z)$ for a Dirac or Weyl semimetal in a magnetic field is shown for both conduction band and valence band states with different Landau level index $N$ [see Eqs.~(\ref{eq:ENWeyl}) and (\ref{eq:E0Weyl})].}
  \label{fig:Landau_cylinders}
\end{figure}

In general, the motion of an electron perpendicular to the magnetic field is quantized into Landau levels, while the electron motion in the field direction is unquantized, so that the Fermi surface breaks up into a set of concentric ``Landau cylinders'' (depicted in Fig.~\ref{fig:Landau_cylinders}a). For electrons in a Dirac or Weyl semimetal, the resulting dispersion relation is
\be 
E_N(k_z) = \text{sign}(N) \sqrt{2 e \hbar B v_x v_y |N| + (\hbar v_z k_z)^2}, \hspace{3mm} (N \neq 0)
\label{eq:ENWeyl}
\ee 
where $N$ is the Landau level index. (Here and below we are neglecting the effect of Zeeman coupling to the electron spin.) The positive-energy Landau levels ($N > 0$) describe conduction band states, while $N < 0$ are valence band states. The level with $N = 0$ has a precisely linear dispersion:
\be 
E_0(k_z) = \pm v_z \hbar |k_z|,
\label{eq:E0Weyl}
\ee 
so that the conduction and valence band levels meet in energy at $k_z = 0$. As we show below, this unusual degeneracy of conduction and valence band Landau levels allows for a huge thermopower at sufficiently large magnetic field.

The semiclassical description of transport works well in the limit where many Landau levels $N$ are populated by conduction band electrons, or in other words the limit where $E_F$ is much larger than the Landau level spacing $\sim \sqrt{e \hbar B v_x v_y}$. As $B$ is increased, the $N > 0$ Landau levels move upward in energy and each Landau level acquires a larger degeneracy. When the magnetic field becomes larger than some value $\beql$, the system enters the ``extreme quantum limit'' (EQL), in which the degeneracy of the lowest Landau level is large enough that all conduction band electrons can reside in the $N=0$ Landau level. In terms of the electron concentration $n$, this field scale is given by
\be 
\beql = \left( \frac{\pi^2}{2 g_0^2} \right)^{1/3} \left( \frac{v_z^2}{v_x v_y} \right)^{2/3} n^{2/3}  \frac{\hbar}{e} 
\label{eq:BEQL}
\ee 
At large fields $B > \beql$, the semiclassical approach fails, and the electron system becomes ``1D-like'' in terms of its dispersion. Only the momentum in the field direction is relevant for the energy of conduction band electrons, so that the Fermi surface for electrons effectively reduces to only two points located at $\kf = \pm \ef/(\hbar v_z)$.

Let us now discuss the behavior of the Seebeck coefficient in each of the three field regimes, $B \ll B_1$, $B_1 \ll B \ll \beql$, and $B \gg \beql$, respectively.

When $B \ll B_1$, the transverse response of the electron system is weak, and the Seebeck coefficient is very weakly affected by the magnetic field. That is, the value of $S_{xx}$ is essentially identical to what is discussed in Sec.~\ref{sec:WeylSxxB0}. (The value of $S_{xx}$ receives a small correction of order $(B/B_1)^2$.)

At intermediate fields $B_1 \ll B \ll \beql$, the Hall response of electrons becomes dominant over the longitudinal response, so that, for example, $\sigma_{xy} \gg \sigma_{xx}$ and $\alpha_{xy} \gg \alpha_{xx}$. Consequently, the Seebeck coefficient is dominated by the Hall conductivity, $S_{xx} \simeq \alpha_{xy}/\sigma_{xy}$ [see Eq.~(\ref{eq:Seebeckdeff})]. However, since both $\alpha_{xy}$ and $\sigma_{xy}$ are proportional to $1/B$, the value of $S_{xx}$ settles into a field-independent constant that is generally of the same order as its value at $B=0$. The precise value of this constant (e.g., the degree of field-enhancement that can be expected by a field of order $B_1$) depends on the dependence of the scattering time $\tau$ on the electron energy. For example, under the assumption of an energy-independent scattering time, $\tau = \textrm{const.}$, the Seebeck coefficient is enhanced by a factor $3/2$ at $B \gg B_1$ (and $B \ll \beql$) as compared to $B = 0$ \cite{kozii_thermoelectric_2019}. Going beyond the semiclassical approximation, one can show that $S_{xx}$ exhibits relatively weak quantum oscillations on top of the semiclassical value. These oscillations are periodic in $1/B$ and are associated with the edges of individual Landau levels passing through the chemical potential.

On the other hand, a field $B \gg \beql$ produces a much stronger and qualitative change to the behavior of the Seebeck coefficient. The origin of this change can be understood via the following simple argument \cite{Obraztsov_1965, jay-gerin_thermoelectric_1974, skinner_large_2018}. Consider a current flowing in the $x$ direction with some drift velocity $v_d$ in the absence of any temperature gradients; we will define the Seebeck coefficient using the ratio of the heat current to the  electric current, $S_{xx} = J^Q_x / (T J^e_x)$ [Eq.~(\ref{eq:SdefOnsager})]. In the limit  of sufficiently large $B$ that $\sigma_{xy} \gg \sigma_{xx}$, the electric field is nearly perpendicular to the flow of current, and in this sense the flow of current is \emph{nearly dissipationless}. Over the course of some time interval $\Delta t$, the electron system transports some amount of heat $\Delta Q = J^Q_x \Delta t$ and some amount of entropy $\Delta \mathcal{S} = J^{\mathcal{S}}_x \Delta t$, where $J^{\mathcal{S}}_x$ is the entropy current density. For any dissipationless process, $\Delta Q = T \Delta \mathcal{S}$, and so we have $J^Q_x \simeq T J^{\mathcal{S}}_x$. The entropy current is given by the entropy density multiplied by the drift velocity $v_d$, while the electric current is given by the charge density multiplied by the same drift velocity, and so we arrive at the following simple and universal equation \cite{ jay-gerin_thermoelectric_1974, skinner_large_2018}:
\be 
S_{xx} = \frac{(\textrm{total electronic entropy})}{(\textrm{net electronic charge})}.
\label{eq:entropypercharge}
\ee 
The only assumption underlying this equation is $\sigma_{xy} \gg \sigma_{xx}$. Thus, in a sufficiently strong magnetic field, the Seebeck coefficient is a direct measurement of the electronic entropy. This relation has been used, for example, to probe the entropy of two-dimensional Quantum Hall systems \cite{Cooper_Halperin_Ruzin_1997, Yang_Halperin_2009, ghahari_kermani_interaction_2014, Sultana_Rienstra_Watanabe_Taniguchi_Stroscio_Zhitenev_Feldman_Ghahari_2025}.

At weak enough fields that $B \ll \beql$, the electronic entropy is proportional to $k_B T/\ef$, and $S_{xx}$ remains small and mostly field-independent. At $B \gg \beql$, on the other hand, the electronic system resides entirely in the zeroth Landau level and the electronic entropy can be enhanced by the magnetic field. The key idea is that the electronic entropy reflects the number of ways that the electrons can be arranged among the available quantum states in the lowest Landau level. Since the degeneracy of the lowest Landau level increases linearly in field, the number of available states also increases linearly in field, and this increase translates to a linear increase in the entropy and therefore the thermopower. A relatively simple calculation gives
\be 
S_{xx} \simeq \frac{g_0}{6} \frac{k_B}{e} \frac{k_B T e B}{\hbar^2 v_z n}.
\label{eq:SxxlinearB}
\ee
This result was first derived in Ref.~\cite{skinner_large_2018}, and the associated linear increase of $S_{xx}$ with $B$ been observed in a number of Dirac and Weyl semimetals in the EQL, including Pb$_{1-x}$Sn$_x$Se \cite{liang_evidence_2013} (with the experiment predating the theory), ZrTe$_5$ \cite{zhang_observation_2020}, TaP \cite{han_tap_2020}, and Bi$_{1-x}$Sb$_x$ \cite{pan_magneto-thermoelectric_2025, he_record_2025} (which used this effect to generate a world record figure of merit $zT \approx 2.6$ at $T = 100$\,K \cite{he_record_2025}). 

\begin{figure}[htbp]
  \centering
  \includegraphics[width=3.0in]{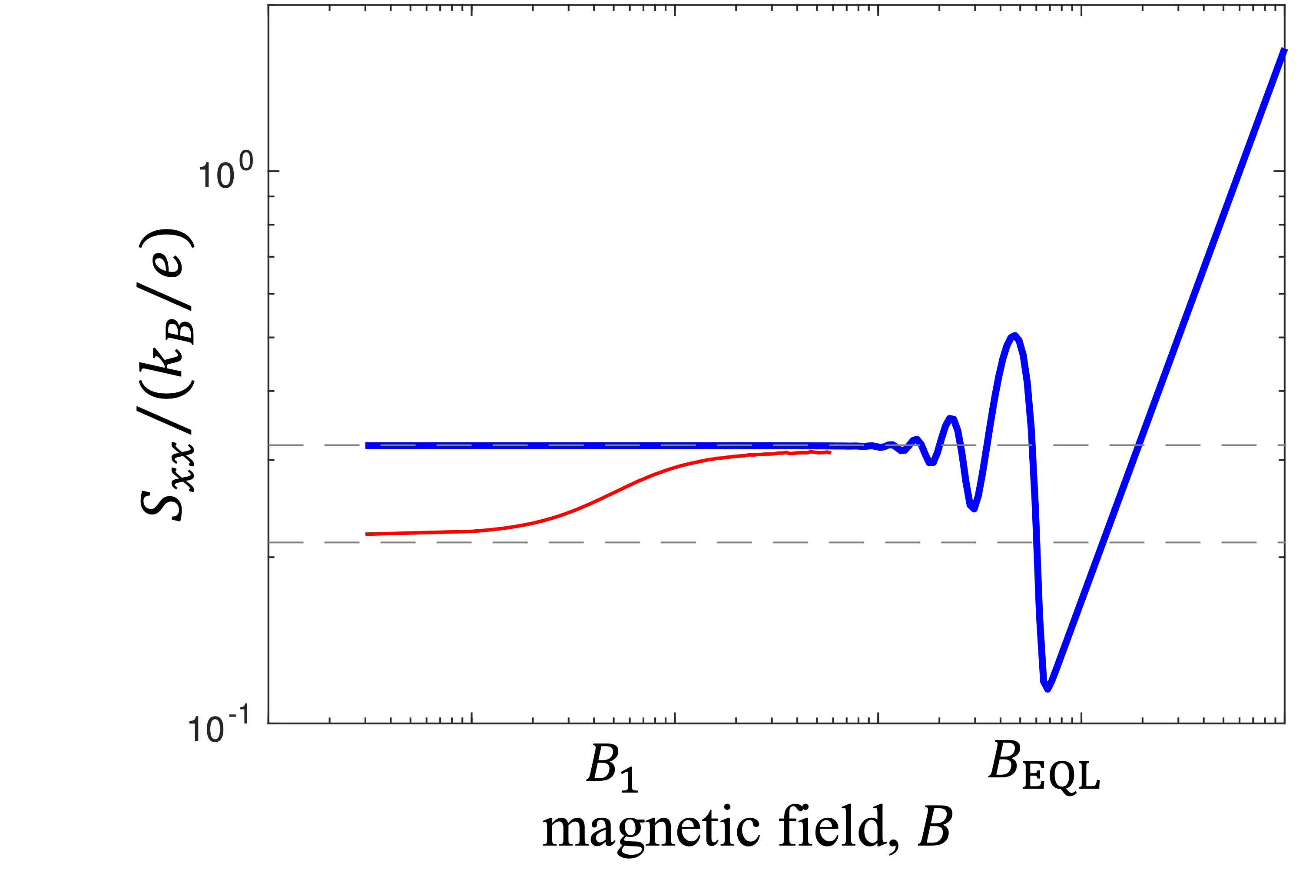}
  \caption{Theoretical calculation of the Seebeck coefficient $S_{xx}$ for a Weyl semimetal with a perpendicular magnetic field. The red curve shows a calculation using the semiclassical approach, which is valid at $B \ll \beql$, while the blue curve shows a calculation using Eq.~(\ref{eq:entropypercharge}), which is valid at $B \gg B_1$. This example uses $k_BT / \ef = 0.03$ and assumes a temperature-independent scattering time. Figure adapted with permission from Ref.~\cite{kozii_thermoelectric_2019}. }
  \label{fig:Sxx-Weyl-B}
\end{figure}

In principle, Eq.~(\ref{eq:SxxlinearB}) implies that the thermopower can be driven arbitrarily high above $k_B/e$ in a conducting system by continually increasing the value of $B$. As the magnetic field is increased significantly above $\beql$, the chemical potential generically falls closer to $E=0$ (since the number of states in the $N=0$ Landau level increases and the electron density is fixed), which for a given temperature implies that an increasing number of holes are excited in the $N=0$ Landau level of the valence band. Nonetheless, Eq.~(\ref{eq:SxxlinearB}) continues to hold even when the chemical potential $\mu \ll k_B T$, so long as the condition $\sigma_{xy} \gg \sigma_{xx}$ is retained, since the thermally excited holes also contribute to the electronic entropy while the total charge (electrons minus holes) remains fixed.

On the other hand, in real materials there are a number of ways that the growth in $S_{xx}$ implied by Eq.~(\ref{eq:SxxlinearB}) can be truncated, so it is worth listing explicitly the assumptions behind its derivation:
\begin{enumerate}
    \item \textbf{\textit{The electron system must be in the extreme quantum limit}}. Since the value of $\beql$ increases with the carrier concentration [Eq.~(\ref{eq:BEQL})], this condition is generally difficult to achieve at realistic  values of $B$ unless the doping is light and carefully controlled. Even if the EQL condition is met at low temperature, increasing the temperature can violate the EQL condition by thermally populating high Landau levels (i.e., the EQL condition is violated if $k_B T \gtrsim \sqrt{\hbar e B v_x v_y}$).

    \item \textbf{\textit{The electron system must be \ three-dimensional}}. While Eq.~(\ref{eq:entropypercharge}) applies equally well to two-dimensional electron systems, in two dimensions the value of $S_{xx}$ does not grow linearly with $B$ because the electrons do not have any entropy associated with free motion along the field direction. Instead the thermopower at $B \gg \beql$ takes the smaller value $\sim (k_B/e) \ln(B/\beql)$.

    \item \textbf{\textit{The chemical potential must remain close to the Weyl point}}. If other electron bands coexist or are nearby in energy to the Weyl or Dirac band, these bands can cause the chemical potential to drift as a function of increasing temperature or magnetic field (as happens, for example, in the Dirac semimetal ZrTe$_5$ \cite{zhang_electronic_2017}). If the chemical potential is drawn away from the Weyl point so that higher Landau levels are populated, then $S_{xx}$ falls to its low-field value.

    \item \textbf{\textit{The electron system must retain large Hall conductivity $\sigma_{xy} \gg \sigma_{xx}$}}. The scattering time $\tau$ of the electron system is in general dependent on both magnetic field $B$ and the temperature. If increasing $B$ or $T$ causes a sufficiently large increase in the electron scattering rate, then the condition of large Hall conductivity can be lost, which causes the value of $S_{xx}$ to revert to its low-field value. 

    \item \textbf{\textit{Opposite-chirality Weyl points must not be too close together}}. Weyl nodes generally come in opposite-chirality pairs (Weyl ``charge'' $\pm 1$), and these pairs have some separation $\Delta k$ in momentum space. If the magnetic field is made so large that $B > e (\Delta k)^2/\hbar$, then the Landau levels of the two Weyl nodes hybridize and form an energy gap \cite{chan_emergence_2017}. These hybridized energy levels no longer exhibit the non-saturating increase in $S_{xx}$. 
\end{enumerate}

Let us assume that these conditions (i)-(v) can be met and consider the behavior of the figure of merit $zT = (S_{xx})^2 T/(\kappa_{xx} \rho_{xx})$. Inserting Eq.~(\ref{eq:SxxlinearB}) gives
\be 
zT \sim \frac{k_B^4 T^3}{\hbar^4} \frac{g_0^2 B^2}{n^2 \kappa_{xx} \rho_{xx} v_z^2}.
\ee 
%
In order to understand what this equation for $zT$ implies about optimal design of Weyl thermoelectrics, one should consider how the resistivity $\rho_{xx}$ depends on the magnetic field and on the Weyl velocities. In the simple semiclassical description, $\rho_{xx} \sim 1/( e^2 v_x^2 \tau g(\mu))$ is independent of field. However, there is no reason that the semiclassical prediction for $\rho_{xx}$ should hold in the EQL, and experiments almost universally observe strong magnetoresistance in the EQL (e.g., Refs.~\cite{murzin_electron_2000, Bhattacharya_Skinner_Khalsa_Suslov_2016, zhang_observation_2020, han_tap_2020, pan_magneto-thermoelectric_2025, he_record_2025}).

For the purpose of commenting on design criteria, let us consider one relatively simple and universal mechanism for producing strong magnetoresistance, which is the mechanism of ``guiding center drift'' in a disorder potential. This mechanism is derived in a semiquantitative way in \ref{sec:guidingcenter}, but at a conceptual level it arises because in a strong magnetic field the dominant mechanism for electron diffusion is associated with electrons undergoing $\vec{E} \times \vec{B}$ drift along contours of constant disorder potential \cite{murzin_electron_2000, song_linear_2015}. Outside the EQL, this mechanism can give rise to linear magnetoresistance. As we discuss in the \ref{sec:guidingcenter}, however, inside the EQL it gives rise to a resistivity that scales as $B^2$ or $B^{7/4}$ (depending on the nature of the disorder potential), driven by the increasing density of states. Using the estimate $\rho_{xx} \sim g_0 e V_0 B^2/(\hbar^2 v_z n^2)$ (see \ref{sec:guidingcenter}), where $V_0$ is the typical electric potential associated with disorder, we arive at the following expression for $zT$ inside the EQL:
\be 
zT \sim \frac{k_B^4 T^3}{e \hbar^2} \frac{g_0}{\kappa_{xx} v_z V_0}, 
\ee 
which suggests that $zT$ may reach a peak or plateau value inside the EQL (as observed in Refs.~\cite{pan_magneto-thermoelectric_2025, he_record_2025}, for example). We emphasize, however, that this peak value is not bound by the same fundamental limitations outlined in Sec.~\ref{sec:conventional}, and it is in principle possible to achieve $zT \gg 1$.

Regardless of which mechanism of magnetoresistance is dominant, one can summarize the following principles for using Weyl semimetals in a magnetic field for thermoelectric applications:

\

\noindent \fbox{%
\parbox{0.96\linewidth}{ \textbf{Optimal design criteria for longitudinal thermoelectric response of Weyl semimetals in a strong magnetic field:}
\begin{itemize}
    \item The carrier concentration $n$ should be low enough that one can reach the EQL, $B > \beql$ [see Eq.~(\ref{eq:BEQL})], at realistic fields.
    \item The velocity $v_z$ in the field direction should be low and the degeneracy $g_0$ of the Weyl point should be high, so that the electronic entropy in the EQL is high for a given temperature.
    \item The velocities $v_x$, $v_y$ in the transport direction should be high, which increases the Landau level spacing (faciliting the EQL) and lowers the resistivity $\rho_{xx}$.
    \item The material should be as little disordered as possible, which reduces $\rho_{xx}$ and weakens magnetoresistance effects. 
     \item The thermal conductivity $\kappa_{xx}$ in the transport direction should be low.
    \item There should be no (or minimal) trivial bands coexisting in energy with the Weyl bands, so that the conductivity in the transport direction is dominated by the Weyl bands and the chemical potential is not pinned to another band with high density of states.
\end{itemize}
} }


\subsection{Nodal-line semimetals}

\begin{figure}[htbp]
  \centering
  \includegraphics[width=3.0in]{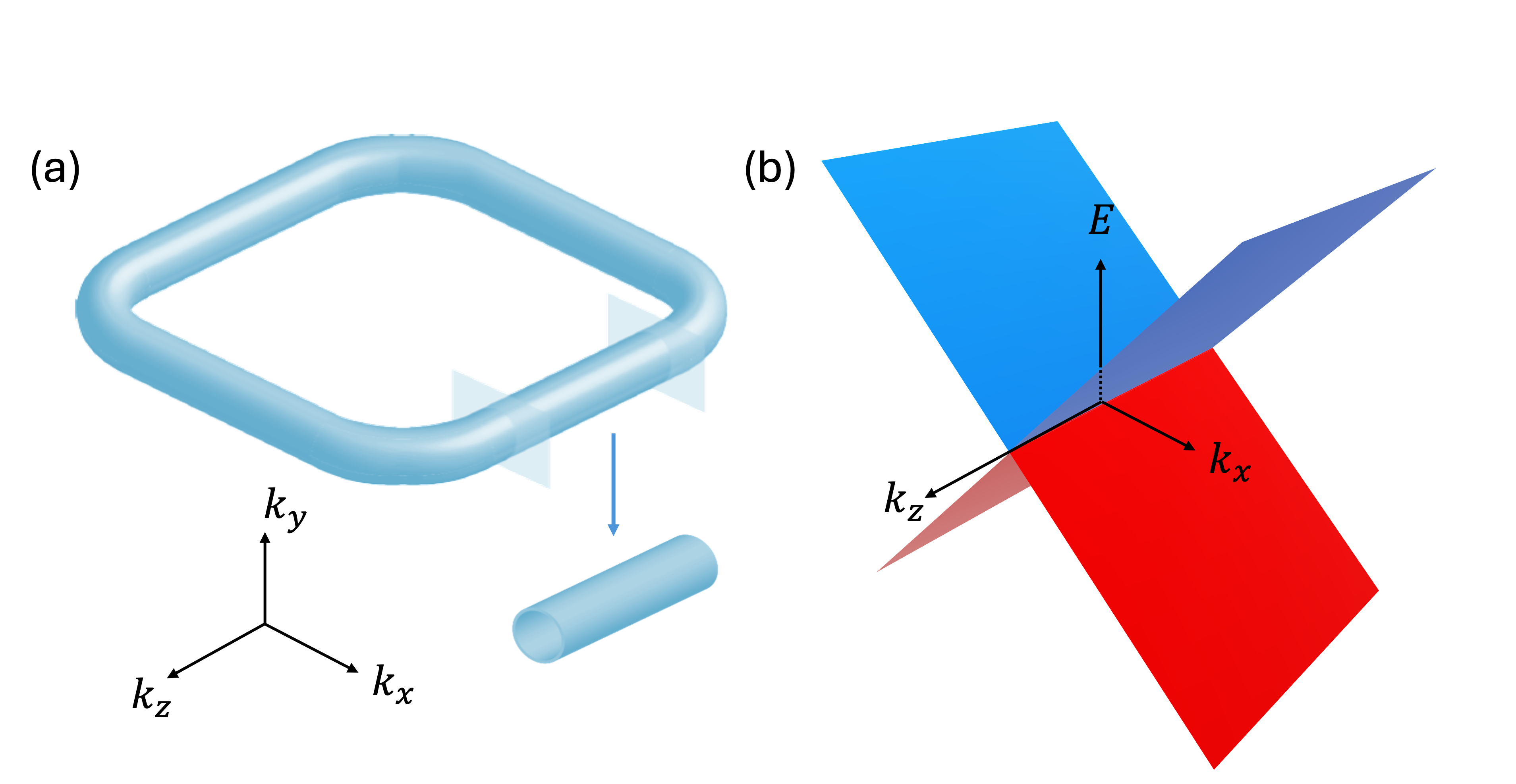}
  \caption{(a) Schematic depiction of the Fermi surface surrounding a nodal line at constant energy, including a short cylindrical segment of the Fermi surface (adapted from Ref.~\cite{syzranov_electron_2017}). (b) The dispersion relation along that segment, showing an energy $E$ that depends linearly on the momentum $k_\perp$ transverse to the nodal line.}
  \label{fig:nodal_line}
\end{figure}

In a nodal-line semimetal, the conduction and valence bands meet at a line or ring in momentum space, and the energy of electrons increases linearly with the momentum $k_\perp$ perpendicular to the nodal line \cite{Burkov_Hook_Balents_2011} (see Fig.~\ref{fig:nodal_line}). In the simplest theoretical models, the nodal line resides at constant energy, so that the density of states is zero at the nodal energy $E=0$ and increases linearly with energy as
\be 
g(E) = \frac{g_0 K_L}{2\pi} \frac{|E|}{\hbar^2 v^2},
\ee 
where $K_L$ is the length of the nodal line in reciprocal space and $v$ is the velocity perpendicular to the nodal line. Such a constant-energy nodal line is conducive to strong thermoelectric response for a similar reason as in Weyl semimetals (i.e., one can achieve low Fermi energy while still maintaining good electrical conductivity), with the added advantage of a greatly enhanced density of states at low energy. At small but finite Fermi energy, the Fermi surface takes the form of a tube that surrounds the nodal line (see Fig.~\ref{fig:nodal_line}).

Unfortunately, in real materials there is no symmetry reason why the nodal line must reside at constant energy, and consequently the nodal line is often ``corrugated'' in energy. In this case, the Fermi surface at $\ef = 0$ comprises neighboring electron and hole pockets along the nodal line (as has been seen, for example, in the nodal line material ZrSiS and \cite{muller_determination_2020}). Such corrugation leads to a finite ``typical'' value of the Fermi energy and the electron density within these electron and hole pockets even when the chemical potential coincides with the average energy of the nodal line. In this sense nodal line semimetals at low doping are often best described as compensated semimetals. The thermoelectric behavior of such compensated semimetals is described in the following section.

On the other hand, if either the temperature or the Fermi energy is large compared to the ``corrugation'' energy scale, then a theoretical description based on a flat nodal line is accurate. The theory for Seebeck and Nernst response in this case was provided in Ref.~\cite{chakraborty_magnetothermopower_2024} (following earlier work on the electrical transport in Ref.~\cite{syzranov_electron_2017}). Here we briefly summarize its main features.

In the absence of magnetic field, the Seebeck response behaves in a way that is qualitatively similar to a Weyl semimetal: the value of $S_{xx}$ is $S_{xx} \sim (k_B/e)(k_B T / \mu)$ at $k_B T \ll \mu$ and $S_{xx} \sim (k_B/e) \mu/(k_B T)$ at $k_B T \gg \mu$, so that the Seebeck coefficient achieves a maximum of order $k_B/e$ when $\mu$ and $k_B T$ are the same order of magnitude. (Again there is the complication that $\mu$ is temperature dependent at large temperatures, so that the actual temperature dependence is $S_{xx} \propto 1/T^2$ at $k_BT \gg \mu$.)

We can therefore conclude that the criteria for optimal design of nodal-line thermoelectrics at $B = 0$ is similar to the case of Weyl semimetals, with the added condition that the nodal line should be relatively flat:

\

\noindent \fbox{%
\parbox{0.96\columnwidth}{ \textbf{Optimal design criteria for longitudinal thermoelectric response of nodal-line semimetals at $B=0$:}
\begin{itemize}
    \item The nodal line should have a nearly-constant energy, with variation in energy that is $\lesssim k_B T$.
    \item The doping should be such that the typical Fermi energy is comparable to $k_B T$.
    \item The velocity in the transport direction $x$ should be high, so that $\rho_{xx}$ is low.
    \item The thermal conductivity $\kappa_{xx}$ should be low.
    \item There should be no (or minimal) trivial bands coexisting in energy with the nodal line, so that the conductivity in the transport direction is dominated by the nodal line states.
\end{itemize}
} }

\ 

If a magnetic field is applied along the direction of the nodal line (or, for a closed nodal ring, within the plane of the nodal ring), it produces Landau quantization of the conduction and valence band states. As with a Weyl semimetal, there is an electron-hole degenerate $N=0$ Landau level, but due to the flatness of the nodal line along the field direction this zeroth Landau level is effectively nondispersive along the field direction. This nondispersive Landau level leads to an enormous density of states at low energy, and consequently to a very large entropy. The field scale $\beql$ associated with the electron system occupying only the lowest Landau level is
\be 
\beql  = \frac{\ef^2}{2 e \hbar v^2} = \frac{(2 \pi)^2 \hbar n}{e g_0 K_L}.
\label{eq:BEQLnodal}
\ee 
At fields $B \gg \beql$, using Eq.~(\ref{eq:entropypercharge}) gives a Seebeck coefficient that is enormous and temperature-independent:
\be 
S_{xx} = 2 \ln 2 \frac{k_B}{e} \frac{B}{\beql}.
\label{eq:SxxNL}
\ee 
This huge thermopower, even at low temperature, arises due to the very large degeneracy of the lowest Landau level, which for a nodal line is degenerate even with respect to motion in the field direction. Since the lowest Landau-level is electron-hole symmetric, it is effectively half-filled even at $\mu = 0$, and there is a large thermodynamic entropy associated with the number of ways to half-fill the Landau level. (Of course, interaction effects may break this degeneracy, as they do in the 2D quantum Hall effect \cite{zhao_symmetry_2010}, so that Eq.~(\ref{eq:SxxNL}) applies only when the temperature is larger than this interaction scale.) The corresponding value of $zT$ is given by \cite{chakraborty_magnetothermopower_2024}
\be 
zT \simeq 0.36 \frac{ (B/B_\text{EQL})^2 }{(\kappa_{xx}/T \, [\textrm{in W}\textrm{m}^{-1}\textrm{K}^{-2}])(\rho_{xx} \, [\textrm{in $\mu \Omega$cm}])}.
\ee 
We emphasize again that this result assumes a flat nodal line. One can think that corrugation of the nodal line introduces a characteristic carrier density $n$, which pushes the field scale $\beql$ to high values [Eq.~(\ref{eq:BEQLnodal})].

We can now summarize the design criteria for nodal-line semimetals in a magnetic field:

\

\noindent \fbox{%
\parbox{0.96\columnwidth}{ \textbf{Optimal design criteria for longitudinal thermoelectric response of nodal-line semimetals in a strong magnetic field:}
\begin{itemize}
    \item The carrier concentration $n$ should be low enough and the circumference $K_L$ of the nodal line should be long enough that one can reach the EQL, $B > \beql$ [see Eq.~(\ref{eq:BEQLnodal})], at realistic fields. The magnetic field should be applied along the nodal line, or within the plane of the nodal line.
    \item The velocity $v$ perpendicular to the nodal line should be high, which increases the Landau level spacing (faciliting the EQL) and lowers the resistivity $\rho_{xx}$.
    \item The nodal line should have a nearly-constant energy, with variation in energy that is significantly smaller than either $k_B T$ or the Landau level spacing $\sqrt{2 e \hbar B v^2}$.
    \item The material should be as little disordered as possible, which reduces $\rho_{xx}$ and weakens magnetoresistance effects. 
     \item The thermal conductivity $\kappa_{xx}$ in the transport direction should be low.
    \item There should be no (or minimal) trivial bands coexisting in energy with the nodal line, so that the conductivity in the transport direction is dominated by the nodal bands and the chemical potential is not pinned to another band with high density of states.
\end{itemize}
} }

\

\subsection{Compensated semimetals (including Type-II nodal semimetals)}
\label{sec:compensated}

In a traditional (non-topological) semimetal, there are both electron- and hole-type bands that intersect the Fermi level at different regions of the Brillouin zone. Consequently the material has some concentration $n_e$ of electrons and $n_h$ of holes simultaneously, so that electrical and thermal currents are carried simultaneously by electrons and holes. Topological semimetals may also have this features if one of the band velocities is \emph{negative}, so that the conduction band dips below the chemical potential $\mu$ and the valence band rises above it even when the nodal point or line is precisely at $E=\mu$. These so-called ``type-II'' nodal semimetals can therefore be thought of as belonging to the class of ``traditional'' semimetals, with the added feature that the electron and hole pockets meet at a point of line in momentum space.

If either $n_e$ or $n_h$ is much larger than the other, then such a semimetal can be expected to behave similarly to a typical single-carrier metal. However, if $\Delta n = n_e - n_h$ is much smaller in magnitude than either $n_e$ or $n_h$, then the material can be described as \emph{nearly completely compensated}, and $\Delta n / n_e$ is a small parameter. In this subsection we focus on this situation, which describes a range of conventional semimetals (such as elemental bismuth and graphite) as well as type-II Weyl and nodal line semimetals (such as WTe$_2$ and Mg$_3$Bi$_2$) and a variety of type-I Weyl semimetals for which there are Weyl points both above and below the Fermi energy (such as TaP \cite{xu_experimental_2015} and SrSi$_2$ \cite{huang_new_2016}).

Such compensated semimetals are generally not considered good candidates for achieving a large Seebeck effect due to the cancellation of the electron and hole contributions. To see this cancellation, one can again think about the Seebeck coefficient as the ratio $J^Q/(T J^e)$ [Eq.~(\ref{eq:SdefOnsager})]. Under the influence of an electric field (and without an applied $B$-field), electrons and holes move in opposite directions, thereby carrying heat in opposite directions (cancelling in $J^Q$) and electric current in the same direction (adding together in $J^e$), so that the Seebeck coefficient is small. For semimetals with large pockets of carriers, the typical Fermi energy $\ef$ within the electron and hole pockets may also be large, so that the heat current is additionally suppressed by a factor $k_B T / \ef$.

\begin{figure}[htbp]
  \centering
  \includegraphics[width=3.0in]{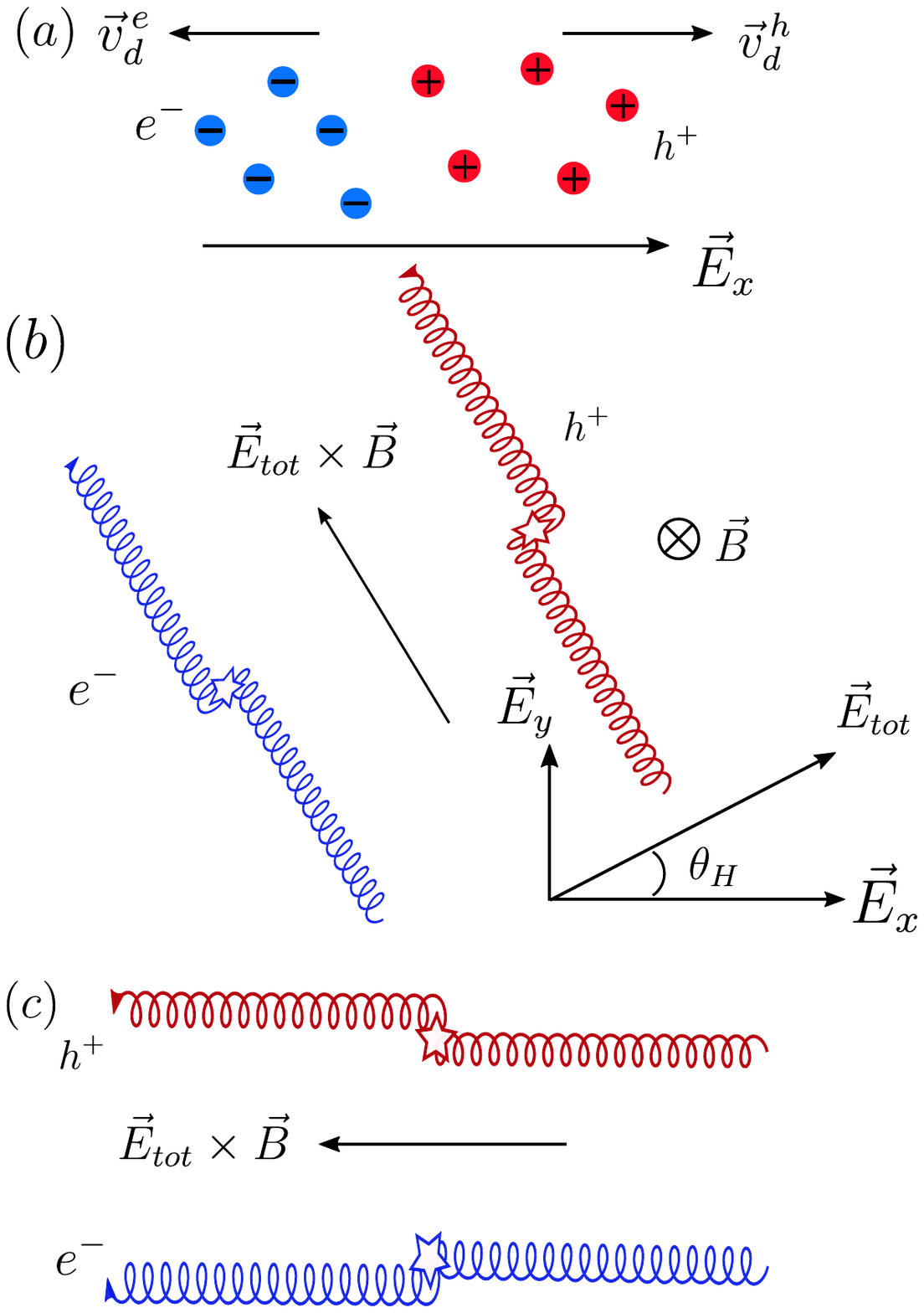}
  \caption{Schematic depiction of the semiclassical trajectories of electrons and holes in a nearly-completely-compensated semimetal ($\Delta n /n_e \ll 1$) at different values of magnetic field. (a) At $B \ll B_1$, electrons and holes drift either with or against the electric field, carrying heat in opposite directions. (b) At $B_1 \ll B \ll B_H$, The electric field has a small transverse component $E_y$, so that the $\vec{E}\times\vec{B}$ drift of carriers has an $x$ component that contributes to the heat current. Star symbols represent instants of electron scattering. (c) At $B \gg B_H$, the Hall field is much larger than the longitudinal field, so that the $\vec{E}\times\vec{B}$ drift is primarily in the $x$ direction and both electrons and holes carry heat in the same direction. Figure taken with permission from Ref.~\cite{feng_large_2021}.}
  \label{fig:e-h-drifting}
\end{figure}

Nonetheless, compensated semimetals can display very strong enhancement of the Seebeck coefficient in a strong magnetic field. The basic reason is that under a sufficiently strong magnetic field that $\omega_c \tau \gg 1$, the motion of electrons and holes is dominated by the $\vec{E} \times \vec{B}$ drift, which results in the drift velocity being nearly parallel for both carrier types (see Fig.~\ref{fig:e-h-drifting}). In this way the B-field allows for the two carrier types to contribute \emph{additively} to the heat current (and even to cancel in the electric current if the field is strong enough).

The theory of $B$-field-enhanced thermopower in compensated semimetals is presented in Ref.~\cite{feng_large_2021}. Here we briefly recap its main features. At $\Delta n / n_e \ll 1$, there are three intrinsic scales of magnetic field: the field $B_1 = \ef/(e \tau v^2)$ associated with the condition $\omega_c \tau = 1$, where $\ef$ is the typical value of the Fermi energy within electron and hole pockets and $v$ is the typical band velocity at the Fermi energy; the field $B_H = B_1 n_e/\Delta n$ associated with developing a large Hall angle ($\sigma_{xy} > \sigma_{xx}$), which is much larger than $B_1$ due to the cancellation of electrons and holes in the Hall effect; and the field $\beql \sim n_e^{2/3} \hbar/e$ associated with the extreme quantum limit. 

At temperatures such that $k_B T \lesssim \ef$, the behavior of the Seebeck coefficient can be described by \cite{feng_large_2021}
\be 
\frac{S_{xx}}{k_B/e} \sim \frac{k_BT}{\ef} \times \begin{cases}
 \Delta n/n_e, & B \ll B_1 \\
 (\Delta n / n_e) (B/B_H)^2, & B_1 \ll B \ll B_H \\
 n_e/\Delta n, & B_H \ll B \ll \beql \\
 (n_e / \Delta n)(B/\beql), & B \gg \beql.
\end{cases}
\label{eq:Sxxcompensated}
\ee 
Notice, in particular, that the Seebeck coefficient experiences a strong, $B^2$ enhancement starting at the relatively low fields $B \sim B_1$ associated with $\omega_c \tau \sim 1$, and its value can be driven to be parametrically larger than $k_B/e$. This large enhancement arises due to the emergence of an $\vec{E}\times\vec{B}$-driven mechanism for carrying heat current: at $B \gg B_1$, electrons and holes drift in a direction that is predominantly perpendicular to the net electric field, so that as long as there is a finite Hall electric field there is a component of the $\vec{E}\times\vec{B}$ drift along the current direction, which produces a significant heat current that is \emph{additive} between electrons and holes (see Fig.~\ref{fig:e-h-drifting}b). At very large fields $B \gg \beql$, there is a linear increase of $S_{xx}$ with $B$ that is similar to Eq.~(\ref{eq:SxxlinearB}), but with an additional large factor $n_e/\Delta n$. (Ref.~\cite{leahy_refining_2025} showed that the sensitive dependence of the thermopower on the ratio $\Delta n / n_e$ can be used to extract the coefficients of a two-band model with much greater sensitivity than the usual fits to longitudinal and Hall resistance.)

Equation (\ref{eq:Sxxcompensated}) suggests that compensated semimetals offer, in principle, the promise of extremely large Seebeck coefficients at sufficiently large fields. In practice, however, such semimetals are unlikely to be more advantageous for Seebeck-based thermoelectric devices than single-band semimetals. One reason is that the magnetic field scales required to achieve large $S_{xx}$ can be large; according to Eq.~(\ref{eq:Sxxcompensated}), achieving $S_{xx} \gtrsim k_B/e$ requires $B \gtrsim B_1 \sqrt{(n_e/\Delta n)(\ef/k_B T)}$. Achieving significant Seebeck coefficient therefore requires high mobility (small $B_1$) and relatively low carrier density for both electron and hole pockets. 
At the same time, nearly-compensated semimetals tend to exhibit huge magnetoresistance effects. Indeed, if the electron and hole bands have the same concentration and same mobility, then $\rho_{xx}(B) = \rho_{xx}(0) (1 + (B/B_1)^2)$ \cite{Ashcroft_Mermin_1976}, so that a high-mobility semimetal can exhibit a huge thermopower that prevents one from achieving large $zT$. Such huge, quadratic magnetoresistance has been, for example, in bulk WTe$_2$, where $\rho_{xx}$ is seen to be enhanced more than $10^5$ times in a field of $60$\,T.

For these reasons, compensated semimetals are generally not good candidates for longitudinal thermoelectrics. As we show below, however, near-compensation can lead to a very strong Nernst response, especially when the typical Fermi energy is small.

\subsection{Topological Insulators}

A three-dimensional topological insulator (TI) is a band-gap insulator that has a gapless, two-dimensional Dirac-type surface state at each of its surfaces. Whether these surface states play a significant role for thermoelectric properties depends on the ratio of the surface to bulk electrical conductance. If the sample is thin enough that the surface states (which may have high mobility relative to the bulk) dominate the conductance, then the thermopower primarily reflects the thermoelectric properties of the surface states.  In this case the thermoelectric properties are similar to those of graphene, for example (see, e.g., \cite{checkelsky_thermopower_2009, hwang_theory_2009}). (TI surface states have the additional feature, not present in graphene, that exact backscattering of electrons is forbidden due to spin-momentum locking. However, since scattering at any other angle is still allowed, the effect of this spin-momentum locking is only to increase the mobility by a factor $\leq 2$.)

On the other hand, if the TI is thick enough, the bulk states inevitably dominate the thermoelectric properties, and the TI behaves similarly to a conventional band gap insulator, so that there are no explicitly ``topological'' features in the thermopower. Of course, bulk TI materials may still demonstrate large Seebeck effect. For example, if the material is poised near the topological-to-trivial transition, the bulk band gap becomes small, and the material becomes a Dirac semimetal precisely at the transition (such a situation seems to arise, e.g., in ZrTe$_5$ \cite{weng_transition-metal_2014, zhang_observation_2020} and in Bi$_{1-x}$Sb$_x$ \cite{vu_thermal_2021}). The bulk band inversion that produces TI may also produce a ``warped'' or flattened bulk band structure, which is conducive to large Seebeck effect because of the enhanced density of states near the band edge. This effect is discussed more completely in Refs.~\cite{he_topology_2025}.

\subsection{Summary of experimental results for magneto-Seebeck effect in topological semimetals, and a word of caution about adiabatic versus isothermal measurements of $S_{xx}$}
\label{sec:Sxxexperiments}

In Table \ref{table:Sxxexp} we give an overview of current experimental results for the Seebeck and magneto-Seebeck effect in nodal semimetals. Particularly notable in terms of the magnitude of the Seebeck effect are the Dirac semimetals ZrTe$_5$ and KMgBi, the Weyl semimetal Bi$_{88}$Sb$_{12}$, and the Weyl semimetals in the TaAs family (TaAs, NbAs, TaP, and NbP), all of which exhibit a large Seebeck coefficient with a huge field enhancement effect. In this section we briefly discuss each of these materials before giving a word of caution about the experimental method by which the Seebeck coefficient is measured.

ZrTe$_5$ is a Dirac semimetal (with, potentially, a small energy gap $< 5$\,meV \cite{zhang_electronic_2017}), for which samples can be grown with very high mobility and low carrier concentration, so that the EQL is achieved already at $B \gtrsim 1.5$\,T \cite{zhang_observation_2020}. The linear-in-$B$ growth of $S_{xx}$ predicted in Eq.~(\ref{eq:SxxlinearB}) is unfortunately truncated at larger $B$ and $T$ by the presence of a coexisting trivial hole band, which causes the chemical potential to sink into the valence band at sufficiently large $B$ and $T$ \cite{zhang_electronic_2017}. Nonetheless, the theoretical result of Eq.~(\ref{eq:SxxlinearB}) describes the data well at $T \lesssim 40$\,K and $B \lesssim 10$\,T \cite{zhang_observation_2020}.

KMgBi is a layered Dirac semimetal with a huge anisotropy of the Dirac velocity: the Dirac velocity is relatively high along the in-plane direction ($\approx 7 \times 10^5$\,m/s) and ``almost flat'' ($\approx 6\times10^3$\,m/s) along the layering direction. Since the Dirac point nominally resides at the Fermi level and is not obscured by any other bands, the material represents a seemingly ideal candidate for field-enhanced Seebeck effect. And indeed, Ref.~\cite{ochs_synergizing_2024} reports a huge Seebeck $S_{xx} \sim 600$\,$\mu$V/K at a field of only $1.5$\,T. Unfortunately, the material is highly air sensitive, which limits its utility for practical applications. 

The bismuth-antimony alloy Bi$_{88}$Sb$_{12}$ is, in some sense, the ideal Weyl semimetal. The substitution of Bi with Sb induces a trivial-insulator-to-topological-insulator transition, so that at a critical value of the doping the bulk band is a gapless Dirac semimetal. Applying a magnetic field splits the Dirac point into two opposite-chirality Weyl points \cite{vu_thermal_2021}. The material can be grown with extremely high purity (with an electron concentration below $10^{16}$\,cm$^{-3}$) and high mobility (exceeding $10^6$\,cm$^{2}$V$^{-1}$s$^{-1}$). 
Within the Weyl semimetal phase, Bi$_{88}$Sb$_{12}$ exhibits a field-enhanced Seebeck coefficient that follows the predictions of Ref.~\cite{skinner_large_2018} [including Eq.~(\ref{eq:SxxlinearB})] almost perfectly within the regime where the temperature and the Landau level spacing are small compared to the energy at which the two Weyl cones merge. While the observed magnitude of $S_{xx}$ is not among the very largest in Table \ref{table:Sxxexp}, the high mobility (low resistivity) allows the material to achieve a record low-temperature figure of merit $zT \approx 2.6$ \cite{he_record_2025} at a magnetic field of only $0.4$\,T, which is well within the range of permanent magnets, and a temperature $100$\,K. (Ref.~\cite{pan_magneto-thermoelectric_2025} reports a similar result of $zT = 1.7$ at $B = 0.7$\,T and $T = 180$\,K.) Further increase of $B$ causes the Seebeck coefficient to increase, but unfortunately this increase is blunted in terms of $zT$ by a large magnetoresistance effects.  Nevertheless, the value of $zT$ exceeds $zT = 1$ for a wide range of temperatures and fields, $50\textrm{ K} < T < 300\textrm{ K}$ and $0.3\textrm{ T} < B < 2\textrm{ T}$ \cite{he_record_2025, pan_magneto-thermoelectric_2025}.

\begin{table*}[!ht]
\centering
 \begin{tabular}{||c c c c c c||} 
 \hline
 \textbf{Material} & \textbf{Type} & \makecell{\textbf{Largest Reported} \\ $|S_{xx}|$ ($\mu$V/K)} & \makecell{\textbf{Conditions} \\ ($T, B$)} & \makecell{\textbf{Field Enhancement} \\ $S_{xx}(T,B)/S_{xx}(T,0)$} & \textbf{Ref.} \\ [0.2ex] 
 \hline\hline 
 Cd\textsubscript{3}As\textsubscript{2} & Dirac & 132 
 & 350 K, 14 T & 2.1 & \cite{wang_cd3as2_2019} \\ \hline
 Cd\textsubscript{3}As\textsubscript{2} & Dirac  & 125 
 & 350 K, 9 T & 5 & \cite{xiang_cd3as2_2020}\\ \hline
 Cd\textsubscript{3}As\textsubscript{2} & Dirac & 75 & 200 K, 13 T & 1.1 & \cite{liang_cd3as2_2017} \\ 
 \hline
 Cd\textsubscript{3}As\textsubscript{2} & Dirac & 650 
 & 5 K, 9 T & 1.4 & \cite{ouyang_cd3as2_2024} \\ \hline 
 CoAs\textsubscript{3} & Dirac & 150 
 & 250 K, 0 T & -- & \cite{sharp_thermoelectric_1995_coas3etc} \\ \hline 
 Fe\textsubscript{3}Sn\textsubscript{2} & Dirac & 22 & 300 K,  0 T & -- & \cite{zhang_fe3sn2_2021} \\ \hline
 IrSb\textsubscript{3} & Dirac & 145 & 750 K,  0 T & -- & \cite{sharp_thermoelectric_1995_coas3etc} \\ \hline
 KMgBi & Dirac & 560 & 80 K, 1.5 T  &  5.6 & \cite{ochs_synergizing_2024} \\ \hline
 NbSb\textsubscript{2} & Dirac & 20 & 25 K, 9 T  &  4.4 & \cite{li_nbsb2_2022} \\ \hline
NbSb\textsubscript{2} & Dirac & 30 & 40 K, 9 T & 7.5 &  \cite{li_nbsb2_2023} \\ \hline
Pb$_{1-x}$Sn$_{x}$Se 
& Dirac & 150 & 44 K, 34 T & 2.1 & \cite{liang_evidence_2013} \\ \hline
PtSn\textsubscript{4} & Dirac & 40 & 11 K, 9 T & 16 & \cite{fu_ptsn4_2020} \\ \hline
RhSb\textsubscript{3} & Dirac & 130 & 350 K, 0 T & -- & \cite{sharp_thermoelectric_1995_coas3etc} \\ \hline
ZrTe\textsubscript{5} & Dirac & 800 & 90 K, 15 T  & 64 & \cite{zhang_observation_2020} \\ \hline
ZrTe\textsubscript{5} & Dirac & 370 & 135 K, 9 T  & 18.5 & \cite{zhang_zrte5_2019} \\ \hline
 Bi$_{1-x}$Sb$_x$ & Weyl & 350 & 80 K, 9 T & 2.9 & \cite{he_record_2025} \\ \hline
 NbAs & Weyl & 450 & 100 K, 14 T & 22.5 & \cite{xu_thermoelectric_2021_taas_tap_nbas_nbp} \\ \hline
NbP & Weyl & 110 
 & 100 K, 9 T & 3.7 & \cite{scott_doping_2023} \\ \hline
 NbP & Weyl & 47 & 110 K, 9 T & 3.1 & \cite{fu_nbp_2018} \\ \hline
NbP & Weyl &   97 
& 100 K, 9 T & 5.7 &  \cite{liu_nbp_2022} \\ \hline 
NbP & Weyl & 1150 & 30 K, 12.5 T & $>$ 23 & \cite{xu_thermoelectric_2021_taas_tap_nbas_nbp} \\ \hline
 TaAs & Weyl & 900 & 35 K, 10 T & 36 & \cite{xu_thermoelectric_2021_taas_tap_nbas_nbp} \\ \hline
 TaP & Weyl & 1600 & 28-40 K, 14 T & 25.6 & \cite{xu_thermoelectric_2021_taas_tap_nbas_nbp} \\ \hline
 TaP & Weyl & 900 & 35 K, 10 T & 18 & \cite{han_tap_2020} \\ \hline
 YPdBi & Weyl & 72 & 300 K, 0 T & -- & \cite{gofryk_ypdbi} \\ \hline
  CoSi & Higher-order Weyl & 92 & 500 K, 0 T & -- & \cite{sk_cosi_2022} \\ \hline
  Co\textsubscript{2}MnAl\textsubscript{1-x}Si\textsubscript{x} & Magnetic Weyl & 40 
 & 300 K, 0 T  & -- &  \cite{Qian_anomalous_2025} \\ \hline
 NdAlSi & Magnetic Weyl & 55 & 350 K, 0 T & -- & \cite{dong_ndalsi_2023} \\ \hline
 TdPtBi & Magnetic Weyl & 250 
 & 150 K, 15 T  & 2.1 &  \cite{wang_tdptbi_2023} \\ \hline
 YbMnBi\textsubscript{2} & Magnetic Weyl & 32 & 304 K, 0 T & -- & \cite{wen_ultrahigh_2025} \\ \hline 
PbTaSe\textsubscript{2} & Nodal Line & 3.5 & 35 K, 9 T & 14 & \cite{yokoi_pbtase2_2021} \\ \hline 
Ta\textsubscript{2}PdSe\textsubscript{6} & Nodal Line & 42 & 20 K, 1 T & $>$ 21 & \cite{nakano_ta2pdse6_2024} \\ \hline
YbMnSb\textsubscript{2} & Nodal Line & 160 & 300 K, 0 T & -- & \cite{pan_ybmnsb2_2021} \\ \hline 
 WTe\textsubscript{2} & Type-II Weyl & 3.5 & 50 K, 0 T & -- & \cite{rana_wte2_2018} \\ \hline
Mg\textsubscript{3}Bi\textsubscript{2} & Type-II Nodal Line & 170 
 & 20 K, 13 T & 17 & \cite{feng_Mg3Bi2_2022} \\ \hline
 \end{tabular}
\caption{An overview of some experimentally reported Seebeck coefficients for topological semimetals. }
\label{table:Sxxexp}
\end{table*}

The TaAs-family Weyl semimetals have type-I Weyl nodes and can be grown with very high mobility, but one should be careful about describing their magnetothermoelectric response via a straightforward application of the theory in Sec.~\ref{sec:WeylSxxB}, since in general they have two inequivalent sets of Weyl points that can be located at different energies \cite{fu_topological_2020}. In their undoped state, the chemical potential in these materials lies in between the two sets of Weyl nodes, so that there are (potentially large) pockets of both electron-type and hole-type carriers \cite{han_tap_2020}. For this reason the magnetothermoelectric response may resemble that of a conventional compensated semimetal (Sec.~\ref{sec:compensated}) \cite{feng_large_2021}. These materials may also exhibit a significant enhancement of thermopower due to phonon drag \cite{xu_thermoelectric_2021_taas_tap_nbas_nbp}.

An important experimental consideration, both for the TaAs-family materials and more generally, is the distinction between \emph{adiabatic} and \emph{isothermal} measurements of the Seebeck effect with an applied magnetic field \cite{harman_thermoelectric_1967}. Adiabatic measurements are the most common experimental method for measuring Seebeck and Nernst effects: they involve using a heater and a heat sink to run a heat current through a material that is otherwise isolated thermally from its environment, and then using thermocouples and voltage probes to measure the resulting temperature and voltage gradients. Notice, however, that in the definition of the Seebeck effect, $S_{xx} = (\Delta V)_x / (\Delta T)_x$, it is assumed that the temperature gradient is purely along the $x$ direction. But this condition may not be achieved, even if the applied heat current is applied along the $x$ direction, due to the thermal Hall effect. That is, the $B$-field-induced transverse motion of carriers leads, generally speaking, to a transverse temperature gradient in response to an applied heat current. If the material exhibits a significant Nernst effect, then this transverse temperature gradient leads to an additional contribution to the longitudinal voltage via the Nernst effect. In this way, the Seebeck signal is polluted (and, potentially, artificially enhanced) by the Nernst effect if there is a significant thermal Hall effect. 

This contamination of the Seebeck signal by the Nernst signal is avoided in an \emph{isothermal} measurement, which involves attaching the sample to a base plate with high thermal conductivity, which prevents the formation of transverse temperature gradients. A careful discussion of the distinction between adiabatic and isothermal measurements is presented in the Supplementary Material of Ref.~\cite{watzman_dirac_2018}. We emphasize here only that when a large field enhancement of $S_{xx}$ is reported, one should check carefully whether the result is not simply a reflection of a large Nernst effect (which, for example, is exhibited by ZrTe$_5$, NbP, TaAs, and TaP, as shown below in Table \ref{table:Sxyexp}) together with a significant thermal Hall effect.

In closing, it is worth commenting on the material YbMnBi$_2$, even though it does not exhibit a particularly large magneto-Seebeck effect in Table \ref{table:Sxxexp}. YbMnBi$_2$ is a type-I Weyl semimetal that in many ways adheres to the ideal criteria for large magneto-Seebeck effect. YbMnBi$_2$ is a layered compound with very small velocity $\approx 6 \times 10^3$\,m/s in the layering direction ($c$) and very high velocity $\approx 1 \times 10^6$\,m/s in the transverse directions (the $a$-$b$ plane) \cite{chaudhuri_optical_2017, borisenko_time-reversal_2019}. In the material's undoped state, the Weyl point resides at the Fermi energy, so that if the doping is light the Fermi surface is small and extremely elongated in the $c$ direction \cite{wen_ultrahigh_2025}. Consequently, the EQL should be easily achievable if the magnetic field is aligned with the $c$ axis, leading to a potentially huge enhancement of the Seebeck coefficient $S_{aa}$ or $S_{bb}$. Existing experiments \cite{pan_ybmnbi2_2022, guo_onsager_2023, wen_ultrahigh_2025} applied the magnetic field in the $a$-$b$ plane, where the EQL is difficult to achieve, and did not observe a significant field enhancement of the Seebeck coefficient (although they did observe a large anomalous Nernst effect, as we discuss below). Exploring the behavior of the Seebeck effect with a $B$-field in the $c$ direction may be a profitable direction for future research (although we caution that the material seems to have a heavy, trivial hole band coexisting with the Weyl point \cite{pan_ybmnbi2_2022, wen_ultrahigh_2025}, which may blunt the growth of $S_{xx}$ with $B$).

Finally, we note that the material CoSi listed in Table \ref{table:Sxxexp} is suspected to be a higher-order topological semimetal, for which four bands meet at a point rather than two \cite{pshenay-severin_band_2018, huber_fermi_2024}. For the purposes of thermopower, such a dispersion should behave similarly to a Dirac or Weyl semimetal.

\section{Transverse (Nernst) Effects}
\label{sec:transverse}

The Nernst effect is the generation of a \emph{transverse} voltage (say, in the $x$ direction) by a temperature gradient (say, in the $y$ direction). The Nernst effect is quantified by the Nernst coefficient $S_{yx} = (\Delta V)_x/(\Delta T)_y$. It is conceptually useful to think about the Nernst coefficient using the Onsager relation with the Peltier effect, in which a transverse heat current is generated by a longitudinal electric current [Eq.~(\ref{eq:SdefOnsager})], $S_{yx} = J^Q_y/(T J_x^e)$, under conditions of uniform temperature.  Like all transverse (linear) responses, the Nernst effect requires the breaking of time reversal symmetry, either through an external applied magnetic field or through the existence of magnetic order (not necessarily ferromagnetic) within the material. 

\begin{figure}[htbp]
  \centering
  \includegraphics[width=3.1in]{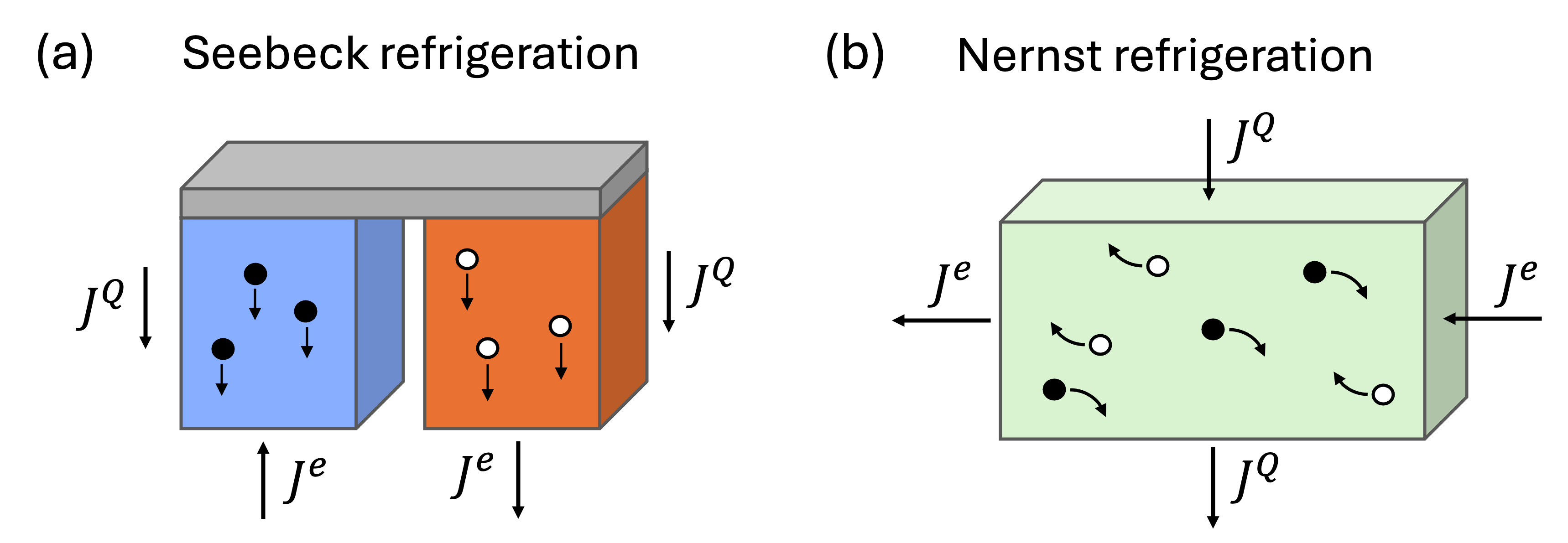}
  \caption{Thermoelectric modules based on (a) the Seebeck effect and (b) the Nernst effect. In a Seebeck-based device, an n-type (blue) and a p-type (red) material are connected in series electrically. When an electric current $J^e$ is run from $n$ to $p$, a heat current $J^Q$ is pulled away from the junction, providing refrigeration. In a Nernst-based device, refrigeration (or heating) of the substrate can be provided by simply running a current through a single material parallel to the substrate, and there is no need for a heterojunction. }
  \label{fig:refrigerators}
\end{figure}

There are a few key conceptual differences between the Seebeck and Nernst effects. At a practical level, the Nernst effect has the advantage that Nerst-based thermoelectric devices do not require a junction between $p$-type and $n$-type materials. For example, a thermoelectric heater or cooler based on the Seebeck effect requires one to make a junction between a $p$-type and an $n$-type conductor, so that heat flows into or out of the junction between them, depending on the direction of the current. For a device based on the Nernst effect, however, one can simply run an electric current through a single material in the direction parallel to a substrate and heat will be removed or deposited into the substrate, depending on the direction of the current (see Fig.~\ref{fig:refrigerators}). For a detailed review of Nernst-based thermoelectric devices we refer the reader to Refs.~\cite{adachi_fundamentals_2025, uchida_thermoelectrics_2022}.

In terms of its origin, a key difference between the Nernst and Seebeck effects is that opposite-sign carriers can contribute additively to the Nernst effect even when the Hall voltage is small or absent. To see this, consider a semimetal with electron and hole bands that are precisely equal in their carrier concentration and mobility. This situation could correspond, for example, to a perfectly compensated conventional semimetal or to a nodal semimetal with the chemical potential precisely at the nodal energy. Such a semimetal must have $S_{xx} = 0$ regardless of the value of temperature or magnetic field (and indeed, by symmetry one cannot argue for either an electron-type or hole-type Seebeck effect). But the Nernst effect is not forbidden by symmetry, since the current direction $\hat{I}$ and the magnetic order (or magnetic field) direction $\hat{M}$ alone establish a unique transverse direction $\hat{I} \times \hat{M}$.

More specifically, suppose in this situation that both the electric current and electric field $E_x$ are in the $x$ direction. Then the electric current density is given by $J^e_x = e (n_e + n_h) v_x$, where $n_e + n_h$ is the combined concentration of electrons and holes and $v_x$ is the drift velocity of carriers in the $x$ direction. The heat current in the $y$ direction can be understood by noting that the flow of charges in the $y$ direction is dissipationless (it is perpendicular to the electric field $E_x$), so one can write $J^Q_y = (\textrm{entropy density})\times T v_y$, where $v_y$ is the drift velocity in the $y$ direction induced by the electric field $E_x$. Then one can write the simple equation
\be 
S_{yx} = \frac{(\textrm{entropy density})}{e (n_e + n_h)} \frac{v_y}{v_x}.
\label{eq:Sxyentropy}
\ee 
This equation is similar to the expression for the Seebeck effect at large $\sigma_{xy}/\sigma_{xx}$ [Eq.~(\ref{eq:entropypercharge})], with a few key differences. First, it corresponds to the limit of \emph{small} $\sigma_{xy}/\sigma_{xx} \ll 1$, rather than to the more difficult-to-achieve limit $\sigma_{xy}/\sigma_{xx} \gg 1$. Second, the denominator includes the combined concentration of electrons and holes, rather than their difference. Third, Eq.~(\ref{eq:Sxyentropy}) depends explicitly on the ratio $v_y/v_x$ of the carrier drift velocities (which for single-band conductors is called the ``tangent of the Hall angle'').

Different mechanisms for generating the Nernst effect differ, essentially, only in the mechanism by which a transverse drift velocity $v_y$ is generated by an electric field. In the remainder of this section, we first consider the behavior of nonmagnetic semimetals in a magnetic field (with a particular focus on nodal semimetals), and then we discuss the behavior of magnetic Weyl semimetals. We refer the reader to the more detailed reviews in Refs.~\cite{watzman_utilizing_2025, adachi_fundamentals_2025} for descriptions of other mechanisms and contexts for the Nernst effect, including extrinsic mechanisms such as skew scattering and side-jump scattering, magnon drag effects, and heterojunction or interface effects.

\subsection{Nonmagnetic semimetals in a magnetic field}

In a nonmagnetic semimetal, including nodal semimetals such as Dirac or Weyl semimetals, there is no Nernst effect without the application of a magnetic field. This is true even for materials with nontrivial Berry curvature -- although the Berry curvature can generate an anomalous velocity for electrons with some particular momentum, the absence of magnetic order guarantees that filled electron states will always have zero Berry curvature on average, so that there is no net transverse velocity. (Nonmagnetic materials can still generate transverse thermoelectric responses at nonlinear order \cite{yang_nonlinear_2025}.) Berry curvature effects are therefore discussed only in the following subsection, when we consider the behavior of magnetic Weyl semimetals.

Let us again assume that the concentrations of electron-type and hole-type carriers are sufficiently similar (and that the $B$-field is weak enough) that $\sigma_{xy} \ll \sigma_{xx}$, so that we can use Eq.~(\ref{eq:Sxyentropy}) to understand optimal design principles for Nernst thermoelectrics.  The transverse drift of carriers $v_y$ is generated by a magnetic field in the $z$ direction (see, e.g., Fig.~\ref{fig:e-h-drifting}b), and the ratio $v_y/v_x$ is generically given by $\omega_c \tau$. The entropy density $s$ is proportional to the typical density of states, $s \sim k_B^2 T g$, where $g$ is given by $g(\mu)$ when $k_B T \ll \ef$ and $g(k_BT)$ when $k_B T \gg \ef$. Consequently we arrive at 
\be 
S_{yx} \sim \frac{k_B}{e} \frac{k_B T g \omega_c \tau}{n_e + n_h}.
\label{eq:Sxyg}
\ee 
Thus, large Nernst effect is associated with high electronic mobility (long scattering time $\tau$) and low combined carrier concentration $n_e + n_h$. For this reason the largest observed Nernst effect so far is in elemental bismuth, a semimetal with high mobility and low carrier concentration, which can exhibit $S_{xy}$ on the order of tens of mV/K (hundreds of times $k_B/e$) \cite{mangez_transport_1976, behnia_nernst_2007}. (Graphite exhibits a similarly large Nernst effect for the same reason \cite{zhu_nernst_2010}.) Notice that if the material is not completely compensated, then at some large enough magnetic field scale $B_H$ a significant Hall voltage develops (see Sec.~\ref{sec:compensated}). At such large fields the Nernst coefficient declines again with magnetic field \cite{feng_large_2021}, so that in general complete compensation of carriers is ideal.

It is worth considering the special case of Eq.~(\ref{eq:Sxyg}) that applies to (nonmagnetic) Weyl or Dirac semimetals. If the chemical potential resides precisely at the nodal point (or, more generally, within $k_BT$ of the nodal energy), then $k_B T g$ and $n_e + n_h$ are of the same order of magnitude, so that one has simply $S_{yx} \sim (k_B/e) \omega_c \tau$. While this equation implies a linear-in-$B$ enhancement of the Nernst effect, the transverse figure of merit [Eq.~(\ref{eq:zTtrans})] depends inversely on the resistance in the $x$ direction, which may also increase significantly with $B$. Assuming sufficiently complete compensation that $\sigma_{xy} \ll \sigma_{xx}$ (or sufficiently small field $B \ll B_H$), there is quadratic magnetoresistance $\rho_{xx} \sim (e^2 g v_x^2 \tau)^{-1} [ 1  + (\omega_c \tau)^2 ]$, so that at $\omega_c \tau \gg 1$ the figure of merit $zT$ reaches a $B$-independent value
\be 
zT \sim \frac{k_B^2 T v_x^2 \tau g}{\kappa_{yy}}.
\label{eq:zTtransnonmag}
\ee 
This result superficially resembles Eq.~(\ref{eq:zTWeylB0}) for the optimal longitudinal $zT$ of nodal semimetals in the absence of a magnetic field (the two equations would be identical if one replaced $\kappa_{yy}$ in Eq.~\ref{eq:zTtransnonmag} with $\kappa_{xx}$). In the former case $zT$ is prevented from ever becoming parametrically larger than zero, since the thermal conductivity $\kappa_{xx}$ can never be smaller than $T v_x^2 \tau g$, which represents the electronic contribution to the longitudinal thermal conductivity. Importantly, however, this limitation does not apply to the transverse $zT$ of Eq.~(\ref{eq:zTtransnonmag}) for two reasons. First, the relevant thermal conductivity is in the direction \emph{transverse} to the electrical current, and $\kappa_{yy}$ can be smaller than $\kappa_{xx}$ (for example, in layered compounds for which the lattice constant is significantly longer in the $y$ direction than in the $x$ direction). In addition, the electron contribution to $\kappa_{yy}$ is generically reduced by the magnetic field, falling as $1/[1 + (\omega_c \tau)^2]$, so that at large $B$ it does not provide a significant constraint to the value of $\kappa_{yy}$.

Since, as mentioned in Sec.~\ref{sec:WeylSxxB0}, most scattering mechanisms have $g \tau = \textrm{const.}$ (by Fermi's golden rule), we conclude that
\be 
(zT)_\textrm{optimal} \propto \frac{v_x^2}{\kappa_{yy}}.
\ee 
One can now summarize the principles of optimal design for transverse thermoelectrics in nodal semimetals as follows:

\

\noindent \fbox{%
\parbox{0.96\linewidth}{ \textbf{Optimal design criteria for transverse thermoelectric response of nonmagnetic nodal semimetals:}
\begin{itemize}
    \item The doping should be such that the Fermi energy is within $k_BT$ of the nodal energy (smaller doping is better).
    \item The magnetic field should be large enough that $\omega_c \tau \gg 1$, but not so large as to produce a large Hall conductivity $\sigma_{xy} \gtrsim \sigma_{xx}$ (i.e., the Hall voltage should be smaller than the longitudinal voltage).
    \item The electron velocity $v_x$ in the transport direction should be high and the lattice thermal conductivity $\kappa_{yy}$ in the transverse direction should be low, so as to maximize the quantity $v_x^2/\kappa_{yy}$.
    \item The material should be as little disordered as possible, which reduces $\rho_{xx}$. 
    \item There should be no (or minimal) trivial bands coexisting in energy with the nodal bands, so that the conductivity (in the transport direction) is dominated by the nodal bands.
\end{itemize}
} }

\

As mentioned above, there is in principle no reason why the value of $zT$ for a transverse thermoelectric cannot become $\gg 1$. 

\subsection{Magnetic Weyl semimetals}
\label{sec:magneticweyl}

In a magnetic material, carriers may have a finite transverse drift velocity $v_y$ even in the absence of magnetic field. In ferromagnets, such a transverse drift can arise from spin-dependent scattering (e.g., skew scattering and side-jump scattering) combined with spin-polarization of the charge carriers. Such spin-dependent scattering processes are responsible for the anomalous Hall effect and anomalous Nernst effect in ferromagnets \cite{nagaosa_anomalous_2010} and are reviewed extensively in Refs.~\cite{adachi_fundamentals_2025, watzman_utilizing_2025}.

\begin{figure}[htbp]
  \centering
  \includegraphics[width=3.0in]{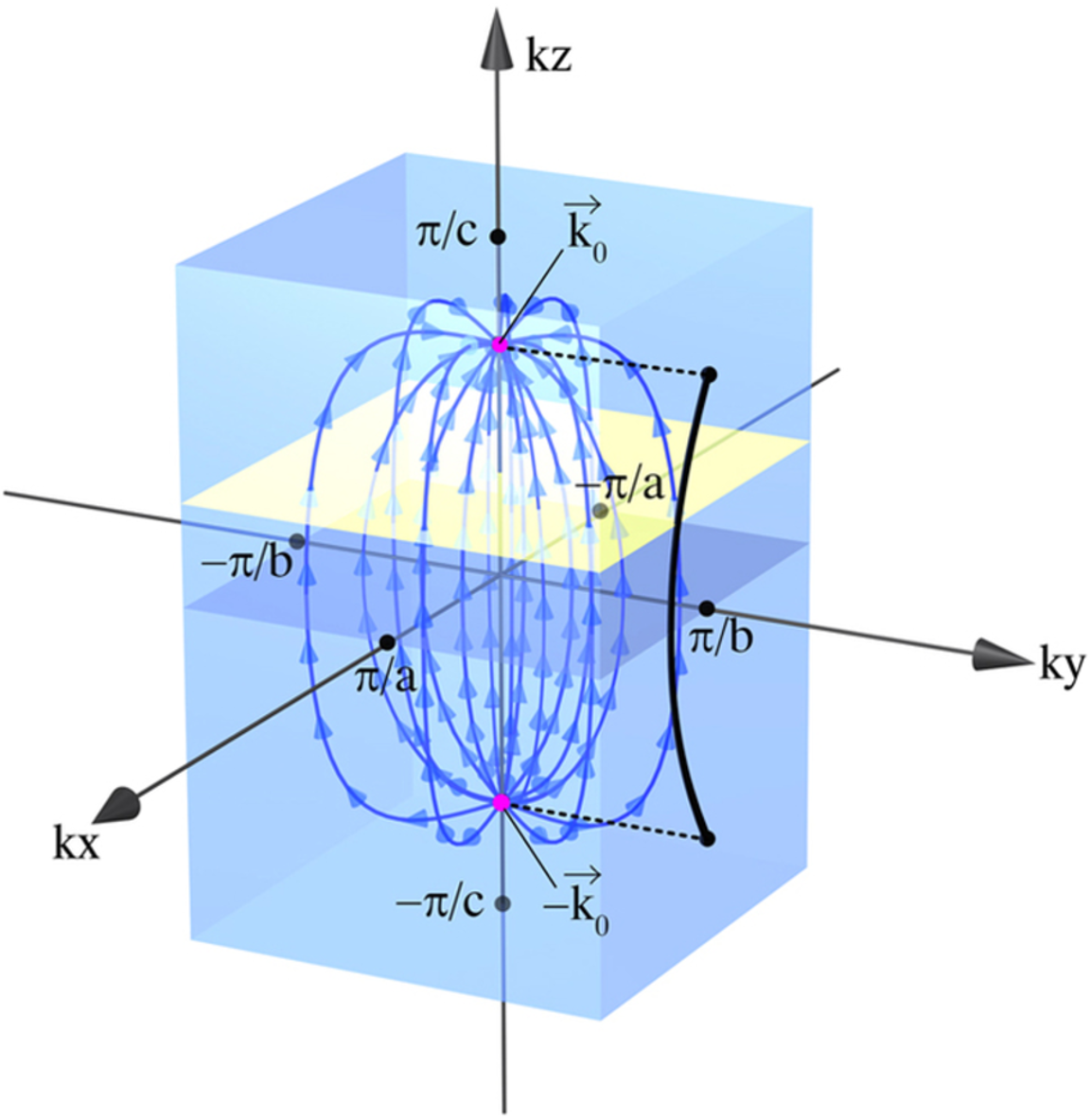}
  \caption{Schematic depiction of the Berry curvature (blue arrows) in a filled band in a magnetic Weyl semimetal. A pair of Weyl points (red dots) with opposite chirality produces a Berry curvature that has a tendency to point in a consistent direction. Figure from Ref.~\cite{morimoto_weyl_2016}. }
  \label{fig:magnetic_weyl}
\end{figure}

In topological materials, the Berry curvature of the band structure provides another mechanism for generating a transverse velocity. Specifically, when electrons experience a Berry curvature $\vec{\Omega}$, an electric field $\vec{E}$ generates an \emph{anomalous velocity} $\vec{v}_A = (e/\hbar) \vec{E} \times \vec{\Omega}$, so that when the electric field is in the $x$ direction electrons experience a transverse drift $v_y$ that is proportional to the component of Berry curvature in the $z$ direction. In this sense the effect of Berry curvature is similar to that of a magnetic field, generating a transverse motion that is analogous to the $\vec{E}\times\vec{B}$ drift. (Unlike the magnetic field, the value of the Berry curvature depends on the momentum of electrons rather than on their position.) Topological band touching points serve as sources of Berry curvature (Weyl points are monopole sources that provide point sources of $\vec{\Omega}$ that are similar to the electric field emanating from a point charge, while nodal lines act like closed flux loops of $\vec{\Omega}$, similar to the magnetic field inside a toroidal solenoid). In non-magnetic materials, the average value of $\vec{\Omega}$ across all filled states at equilibrium is zero. In magnetic materials, however, the average value of $\vec{\Omega}$ across all filled states need not be zero (see Fig.~\ref{fig:magnetic_weyl}), and when this is the case there is an \emph{anomalous Hall conductivity} $\sah$ in the plane perpendicular to the net direction of $\vec{\Omega}$.

Let us now focus our discussion on the properties of magnetic Weyl semimetals, such as Co$_3$Sn$_2$S$_2$ \cite{yang_co3sn2s2_2020} or YbMnBi$_2$ \cite{pan_ybmnbi2_2022, wen_ultrahigh_2025}, in the absence of a magnetic field. In other words, we will consider the \emph{anomalous Nernst effect} in magnetic Weyl semimetals. In these materials, an anomalous Hall conductivity $\sah$ is generated by the filled conduction band states. The value of $\sah$ is maximal when the chemical potential resides precisely at the band touching point and the valence band is completely filled. As the chemical potential is moved away from the nodal point, the value of $\sah$ is reduced, since conduction and valence bands have opposite signs of Berry curvature. For this reason, in contrast to the magnetic field case, there is generally no anomalous Nernst effect at $\mu = 0$, since thermally-excited electrons and holes experience opposite Berry curvature. (Formally, the Nernst effect disappears at $\mu = 0$ since $\sigma_{xy}$ is an even function of the energy $E$, so that the integral expression for the Nernst conductivity in Eq.~(\ref{eq: alpha conductivity}) disappears.) 

Let us assume instead that the chemical potential $\mu$ is finite, and that $\sigma_{xy} = -\sigma_{yx} = \sah$ is an energy-independent constant. This assumption is equivalent to assuming that $\mu$ is much smaller than the band width (i.e., light doping). At $B = 0$, this set of assumptions produces $\alpha_{xy} = 0$, and the Nernst coefficient is given by $S_{yx} = \alpha_{xx} \sah/[\sigma_{xx} \sigma_{yy} + (\sah)^2]$. The value of the longitudinal thermoelectric conductivity $\alpha_{xx}$ is of order $\alpha_{xx} \sim (k_B/e) \sigma_{xx} k_B T / \ef$ when $k_B T \ll \ef$, and it is maximized when the doping is such that $\ef$ and $k_B T$ are of the same order (at smaller values of the doping, thermally excited holes in the valence band cancel the electron contribution to $\alpha_{xx}$). If we assume that $\sah$ is smaller than $\sqrt{\sigma_{xx} \sigma_{yy}}$ (as is so far the case in observations of the anomalous Hall effect in magnetic Weyl semimetals, e.g., Refs.~\cite{wang_large_2018, alam_sign_2023}), then we arrive at
\be 
S_{yx} \sim \frac{k_B^2 T}{e \ef} \frac{\sah}{\sigma_{yy}}.
\label{eq:Syxsah}
\ee 
So achieving a large anomalous Nernst effect requires a large anomalous Hall conductivity and a small electrical conductivity in the $y$ direction, and $\ef \sim k_B T$.

Using Eq.~(\ref{eq:zTtrans}) we can estimate the transverse figure of merit:
\be 
zT \sim \left( \frac{k_B^2 T \sah}{e \ef \sigma_{yy}} \right)^2 \frac{T \sigma_{xx}}{\kappa_{yy}}.
\ee 
If we assume that $k_B T$ and $\ef$ are of the same order of magnitude (which produces the maximal anomalous Nernst coefficient), and inserting expressions for the conductivities in terms of the Weyl velocities, we arrive at
\be 
(zT)_\text{opt} \sim \frac{k_B^2 T (\sah)^2 v_x^2}{e^4 \kappa_{yy} v_y^4 g \tau},
\label{eq:zTmagneticWeyl}
\ee 
where $g$ is the typical density of states, and again $g \tau$ is typically a constant that depends on the disorder concentration and on the temperature but not on material parameters. 

Equation (\ref{eq:zTmagneticWeyl}) allows us to summarize optimal design criteria for Weyl thermoelectrics based on the anomalous Nernst effect:

\ 

\noindent \fbox{%
\parbox{0.96\linewidth}{ \textbf{Optimal design criteria for anomalous transverse thermoelectric response in magnetic Weyl semimetals:}
\begin{itemize}
    \item The doping should be such that the Fermi energy is of order $k_BT$ away from the Weyl point.
    \item The anomalous Hall conductivity $\sah$ should be large (but not so large as to develop a large Hall angle).
    \item The electron velocity $v_x$ in the current direction should be high, the velocity $v_y$ in the transverse direction should be low, and the lattice thermal conductivity $\kappa_{yy}$ in the transverse direction should be low, so as to maximize the quantity $v_x^2/\kappa_{yy} v_y^4$.
    \item There should be no (or minimal) trivial bands coexisting in energy with the nodal bands, so that the conductivity (in the transport direction) is dominated by the Weyl bands.
\end{itemize}
} }

\

Notice, in particular, that optimizing $zT$ in this case does not rely on high mobility. On the contrary, the momentum relaxation time $\tau$ appears in the denominator of Eq.~(\ref{eq:zTmagneticWeyl}), so that, under the assumption of $\sah \ll \sqrt{\sigma_{xx} \sigma_{yy}}$, lower mobility is preferable for achieving higher $zT$. This conclusion arises because higher mobility (longer $\tau$) generally increases the transverse conductivity $\sigma_{yy}$ as well as the longitudinal conductivity, while the anomalous Hall conductivity $\sah$ is generally set by the band structure and is independent of $\tau$.

Of course, in the presence of a magnetic field, a magnetic Weyl semimetal can exhibit large Nernst effect due to the combined effects of both the anomalous contribution [Eq.~(\ref{eq:Syxsah})] and the field-induced contribution [Eq.~(\ref{eq:Sxyg})]. Such a synergistic combination is apparently responsible for the record anomalous Nernst effect $S_{yx}^\text{AH} \sim 70\,\mu$V\,K$^{-1}$ observed in the Weyl antiferromagnet YbMnBi$_2$ \cite{wen_ultrahigh_2025}. This material also exhibits an extremely anisotropic dispersion, $v_x/v_y \sim 200$, so that the mobility in the transport direction can remain high (facilitating the field-induced Nernst effect) even while the transverse conductivity $\sigma_{yy}$ is low (enabling the anomalous Nernst effect). Comparison with previous studies on the same material \cite{pan_ybmnbi2_2022}, which had a higher doping and observed an anomalous Nernst effect that was an order of magnitude smaller, makes clear the crucial role of having a low Fermi energy.


\subsection{Summary of experimental results for Nernst effect in topological semimetals}

\begin{table*}[htb]
\centering
 \begin{tabular}{||c c c c c||} 
 \hline
 \textbf{Material} & \textbf{Type} & \makecell{\textbf{Largest Reported} \\ $|S_{yx}|$ ($\mu$V/K) } & \makecell{\textbf{Conditions} \\ ($T, B$)} & \textbf{Ref.} \\ [0.2ex] 
 \hline\hline
 Cd\textsubscript{3}As\textsubscript{2} & Dirac & 600 & 5 K, 6-9 T & \cite{ouyang_cd3as2_2024} \\ \hline
 Cd\textsubscript{3}As\textsubscript{2} & Dirac & 125 & 350 K, 3 T & \cite{xiang_cd3as2_2020} \\ \hline
 Cd\textsubscript{3}As\textsubscript{2} & Dirac & 10 & 8.5 K , 5 T & \cite{liang_cd3as2_2017} \\ \hline
 Fe\textsubscript{3}Sn\textsubscript{2} & Dirac & 2.1 & 300 K, 1.5 T & \cite{zhang_fe3sn2_2021} \\ \hline
 NbSb\textsubscript{2} & Dirac & 616 & 21 K, 9 T & \cite{li_nbsb2_2022} \\ \hline
 NbSb\textsubscript{2} & Dirac & 400 & 40 K, 9 T & \cite{li_nbsb2_2023} \\ \hline
 PtSn\textsubscript{4} & Dirac & 45 & 6.2 K, 9 T & \cite{fu_ptsn4_2020} \\ \hline
 ZrTe\textsubscript{5} & Dirac & 1200 & 100 K, 15 T & \cite{zhang_observation_2020} \\ \hline
 ZrTe\textsubscript{5} & Dirac & 420 & 145 K, 9 T & \cite{zhang_zrte5_2019} \\ \hline
 Mg\textsubscript{3}Bi\textsubscript{2} & Weyl & 617 & 15 K, 14 T & \cite{feng_mg3bi2_2023_rsc} \\ \hline
 NbAs & Weyl & 500 & 140 K, 14 T & \cite{xu_thermoelectric_2021_taas_tap_nbas_nbp} \\ \hline
 NbP & Weyl & 800 & 109 K, 9 T & \cite{watzman_dirac_2018} \\ \hline
 NbP & Weyl & 800 & 50 K, 9 T & \cite{stockert_nbp_2017} \\ \hline
 NbP & Weyl & 700 & 150 K, 14 T & \cite{xu_thermoelectric_2021_taas_tap_nbas_nbp} \\ \hline
 NbP & Weyl & 124 & 140-300 K, 9 T & \cite{liu_nbp_2022} \\ \hline
 NbP & Weyl & 100 & 163.8 K, 9 T & \cite{scott_doping_2023} \\ \hline
 NbP & Weyl & 80 & 122 K, 9 T & \cite{fu_nbp_2018} \\ \hline
 TaAs & Weyl & 1750 & 60 K, 14 T & \cite{xu_thermoelectric_2021_taas_tap_nbas_nbp} \\ \hline 
 TaP & Weyl & 1750 & 40 K, 14 T & \cite{xu_thermoelectric_2021_taas_tap_nbas_nbp} \\ \hline
 TaP & Weyl & 1000 & 40 K, 9 T & \cite{han_tap_2020} \\ \hline
 WTe\textsubscript{2} & Type II Weyl & 7000 & 12.5 K, 9 T & \cite{pan_wte2_2022} \\ \hline
WTe\textsubscript{2} & Type II Weyl & 35 & 3 K, 9 T & \cite{rana_wte2_2018} \\ \hline
 CoSi & Higher-order Weyl & 180 & 42 K, 15 T & \cite{xu_crystal_2019} \\ \hline
EuMnBi\textsubscript{2} & Magnetic Dirac & 33 & 14.5 K, 8.5 T & \cite{pan_magnetic_topological_sm_thermopower} \\ \hline
 Co\textsubscript{2}MnAl$_{1-x}$Si$_{x}$ & Magnetic Weyl & 4.9 & 300 K, 1 T & \cite{Qian_anomalous_2025} \\ \hline
Co\textsubscript{2}MnGa & Magnetic Weyl & 8 & 400 K, 2 T & \cite{sakai_co2mnga_2018} \\ \hline
Co\textsubscript{2}MnGa & Magnetic Weyl & 6 & 340 K, 1 T  & \cite{guin_co2mnga_2019} \\ \hline
 Co\textsubscript{3}Sn\textsubscript{2}S\textsubscript{2} & Magnetic Weyl & 5$^*$ & 70 K, 0 T & \cite{yang_co3sn2s2_2020} \\ \hline
 EuCd\textsubscript{2}As\textsubscript{2} & Magnetic Weyl & 4.5 & 200 K, 3 T & \cite{roychowdhury_eucd2as2_2023} \\ \hline 
 Mn\textsubscript{3}Sn & Magnetic Weyl & 0.6$^*$ & 200 K, 0 T & \cite{tomita_mn3sn_nodate} \\ \hline
 TdPtBi & Magnetic Weyl & 214 & 300 K, 13.5 T & \cite{wang_tdptbi_2023} \\ \hline 
 YbMnBi\textsubscript{2} & Magnetic Weyl & 110 & 254 K, 5-9 T & \cite{wen_ultrahigh_2025} \\ \hline 
 NbAs\textsubscript{2} & Nodal Line & 600 & 35 K, 9 T  &  \cite{wu_nbas2_2024} \\ \hline
 PbTaSe\textsubscript{2} & Nodal Line & 3 & 17 K, 9 T & \cite{yokoi_pbtase2_2021} \\ \hline
 Ta\textsubscript{2}PdSe\textsubscript{6} & Nodal Line & 150 & 20 K, 4 T  & \cite{nakano_ta2pdse6_2024} \\ \hline
 Mg\textsubscript{3}Bi\textsubscript{2} & Type II Nodal Line & 130 
 & 15 K, 13 T & \cite{feng_Mg3Bi2_2022} \\ \hline
 \end{tabular}
 \caption{An overview of some experimentally reported Nernst coefficients for topological semimetals. Measurements at $B = 0$ (anomalous Nernst effects) are highlighted with an asterisk, although other Magnetic Weyl materials generally also have an anomalous contribution to the Nernst signal.}
 \label{table:Sxyexp}
\end{table*}

Table \ref{table:Sxyexp} presents an overview of experimental results for the Nernst effect (traditional and anomalous) in nodal semimetals. Prominent on this list are the Dirac semimetal ZrTe$_5$ and Weyl semimetals in the TaAs family (TaAs, TaP, NbAs, and NbP), which also appear prominently in Table \ref{table:Sxxexp}. Their appearance on both lists is not surprising, since the features that allow them to have a large Nernst effect -- high mobility and low Fermi energy -- also enable a large magneto-Seebeck effect. Of course, the large Nernst effect means that the large apparent magneto-Seebeck effect should be interpreted with caution, since the Nernst signal can contaminate the Seebeck signal when there is a thermal Hall effect, as we discuss in Sec.~\ref{sec:Sxxexperiments}.

The largest Nernst effect in Table \ref{table:Sxyexp}, however, corresponds to the type-II Weyl material WTe$_2$, which achieves $S_{xy} \sim 7000$\,$\mu$V/K despite having a small Seebeck effect. This combination of large Nernst and small Seebeck effect can be rationalized by the near-perfect compensation of electron and hole carriers in WTe$_2$ combined with high mobility. The high mobility allows the experiments to reach a very large value of $\omega_c \tau$, which drives the Nernst coefficient very high [see Eq.~(\ref{eq:Sxyg})], while the near-complete compensation of electrons and holes prevents the electron system from achieving large Hall angle, so that there is no corresponding enhancement of the Seebeck effect. In this sense there is nothing explicitly ``topological'' about the Nernst effect in WTe$_2$, and one can view the large Nernst effect as arising for a similar reason as in conventional semimetals such as graphite or elemental bismuth. The large Nerst effect allows the transverse figure of merit $zT$ in WTe$_2$ to become as large as $zT \sim 0.3$ (at $T = 12$\,K, $B = 9$\,T), with an accompanying huge power factor \cite{pan_wte2_2022}, despite the large quadratic magnetoresistance.

The nodal line material NbAs$_2$ also achieves a fairly large Nernst effect, and this effect is likely also attributable to conventional semimetal physics, since the corrugation of the nodal line in energy produces relatively large electron and hole pockets \cite{wu_nbas2_2024}.

Included in Table \ref{table:Sxyexp} are two measurements of a $B = 0$ (strictly anomalous) Nernst effect, which are relatively small in magnitude. In principle all magnetic Weyl materials exhibit an anomalous contribution to the Nernst effect, although in the typical measurement of $S_{xy}$ vs.\ $B$ this anomalous contribution can be difficult to separate from the field-induced contribution unless there is strong magnetic hysteresis. For the Co-based magnetic Weyl materials in Table \ref{table:Sxyexp} it is relatively clear that the bulk of the measured $S_{xy}$ can be attributed to the anomalous Hall effect, so that they establish an anomalous Nernst effect as large as 8\,$\mu$V/K (experimental results for the anomalous Nernst effect up to the year 2020 are reviewed nicely in Ref.~\cite{fu_topological_2020}). Ref.~\cite{wen_ultrahigh_2025}, on the other hand, claims a record anomalous Nernst effect of $\sim 70$\,$\mu$V/K, which contributes to the total measured Nernst coefficient $\approx 110$\,$\mu$V/K. This anomalous Nernst effect is explained as arising from a spin canting transition in the material, which splits Dirac points in the band structure into pairs of opposite-chirality Weyl points and turns on the anomalous hall conductivity $\sah$. For this reason a finite magnetic field is required to align the spin canting direction across the material, so that the Nernst effect is absent at $B = 0$.


\section{High-throughput search for optimal topological thermoelectric materials}
\label{sec:search}

In the preceding sections we have summarized the optimal design criteria for a variety of situations involving longitudinal and transverse thermoelectric response. While different situations call for different criteria in terms of the doping and magnetic field (for a given temperature), in terms of materials selection there is a nearly-universal set of five desirable properties for a topological semimetal:
\begin{enumerate}
    \item The nodal energy must reside at (or very near) the Fermi level for the undoped material.

    \item There should be no (or very little) overlap of other bands with the nodal energy.

    \item The velocity $v_x$ in the transport direction should be as large as possible.

    \item The velocity $v_z$ in the magnetic field direction should be as small as possible.

    \item The thermal conductivity ($\kappa_{xx}$ or $\kappa_{yy}$) should be low. 
\end{enumerate}
Only the case of thermoelectrics based on the anomalous Nernst effect (see Sec.~\ref{sec:magneticweyl}) impose any selection criteria outside this list (requiring, in addition, a large anomalous Hall conductivity and a small velocity $v_y$).

For all other cases, if a material is identified that meets these criteria, then the primary practical consideration is whether the doping can be controlled during chemical synthesis with sufficient precision to produce a low Fermi energy (ideally comparable to $k_B T$). If so, then maximizing $zT$ is a matter only of setting the temperature and magnetic field (and its direction) to the optimal values.

One can notice that the criteria (i)-(iv) listed above are properties only of the electronic band structure. Thus, a large-scale search for ideal topological materials can be performed by searching through online databases of electronic band structure calculated using density functional theory (DFT). In this section we present results from such a search, using data available in the Topological Materials Database (TMDB) \cite{vergniory_all_2022, vergniory_complete_2019, bradlyn_topological_2017, aroyo_bilbao_2006} (online at \href{https://topologicalquantumchemistry.org/}{https://topologicalquantumchemistry.org/}).

\begin{figure}[htb!]
  \centering
  \includegraphics[width=3.0in]{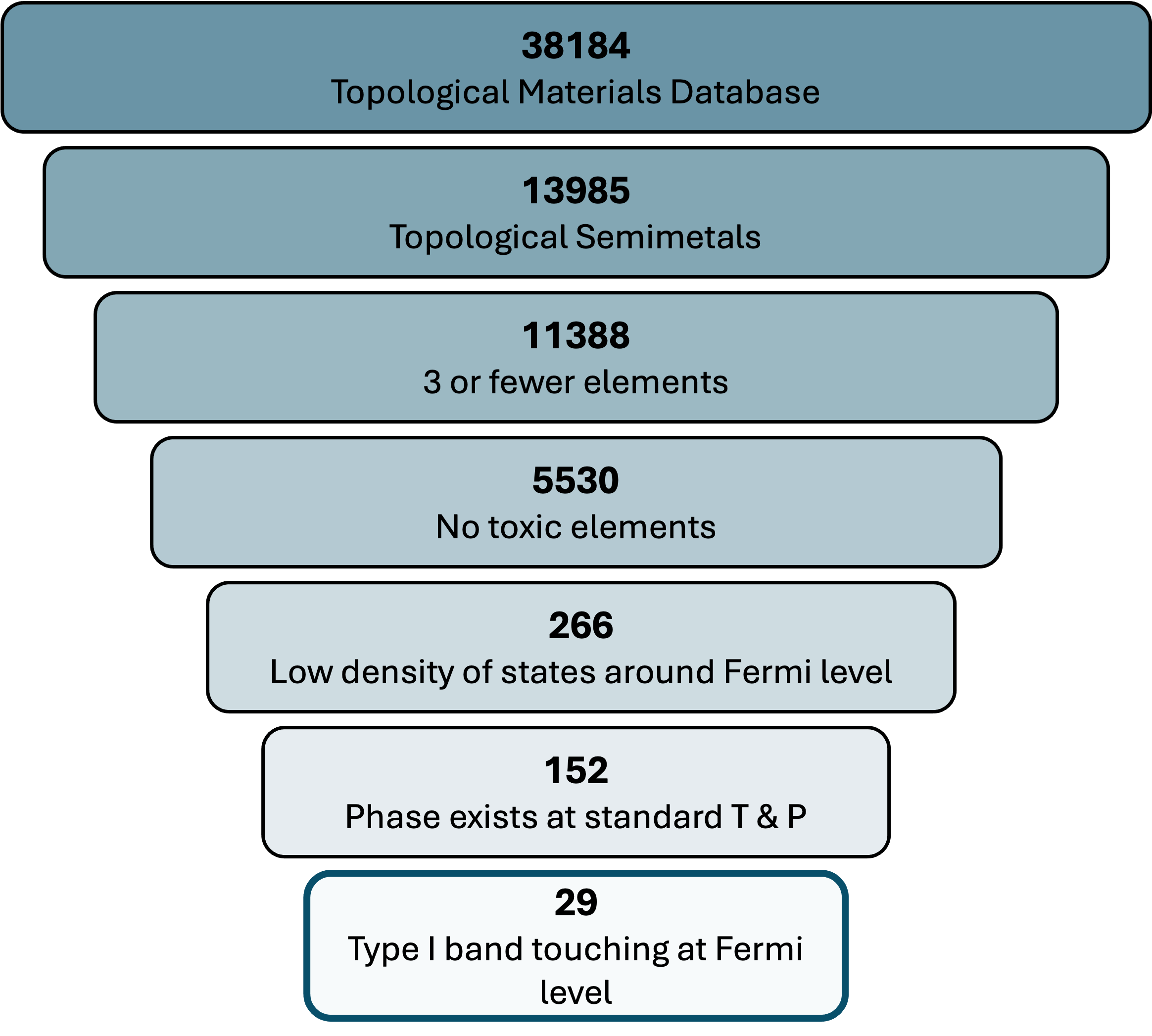}
  \caption{Illustration of the filtering of materials in our computational search for topological thermoelectrics. 38,184 entries from the TMDB are filtered to produce a list of 29 candidates.}
  \label{fig:flowchart}
\end{figure}

The process of our computational search is summarized in Fig.~\ref{fig:flowchart}. Briefly, we consider all 13,985 crystalline materials that are labeled as topological semimetals in the TMDB. In order to focus on materials that are realistic to synthesize, we discard compounds that have more than three distinct elements, crystal structures that do not exist at standard temperature and pressure, and compounds that include any of the elements U, Os, Rb, Cs, Hg, Tl, Tc, Po, Xe, Be, or Th, which are toxic and/or radioactive. Our goal is to identify materials that have a point or line node at the Fermi energy ($E = 0$) and that have no other bands coinciding with the Fermi energy. Such materials have zero density of states at $E = 0$, but since the density of states data in the TMDB is often smoothed over a finite energy window, we use a conservative approximation for this criterion by filtering out any materials for which the density of states exceeds $1$\,eV$^{-1}$\,\AA$^{-3}$ over the entire window $-0.5\textrm{ eV} < E < 0.5\textrm{ eV}$. For the 152 compounds that remain after this filtering, we analyze the dispersion relation along the path in reciprocal space (``$k$-path'') provided in the TMDB in order to filter out any materials that do not exhibit a type-I band touching point near $E=0$ along the $k$-path.  We arrive at a list of 29 compounds, which include 27 point-node materials and two line-node materials. For each of these materials we calculate the band velocity $(1/\hbar) dE/dk$ near the band touching point for each crystallographic direction provided in the $k$-path. Our results are shown in Table \ref{table:searchresults} (for point nodes) and Table \ref{table:nodalsearchresults} (for line nodes).

\begin{table*}[!ht]
\centering
 \begin{tabular}{||c c c c c||} 
 \hline
 \textbf{Material} & \textbf{ICSD} & \textbf{Space Group} & \makecell{\textbf{Max Velocity} (m/s) \\ (direction, band) } & \makecell{\textbf{Min Velocity} (m/s) \\ (direction, band) } \\ [0.2ex] 
 \hline\hline
        AgAsSr & 49742 & P6$_3$/mmc & $2.1\times10^5$ ($\Gamma$ to A, CB) & $1.6\times10^4$ ($\Gamma$ to A, VB) \\ \hline
        AlY$_3$C & 43869 & Pm$\bar{3}m$ & $6.7\times10^4$ ($\Gamma$ to R, CB) & $2.4\times10^4$ (M to $\Gamma$, VB) \\ \hline
        Ba$_6$Ga$_5$N & 77731 & R$\bar{3}c$ & $2.4\times10^5$ (T to $\Gamma$, VB) & $5.0\times10^4$ ($\Gamma$ to L, CB) \\ \hline
        BaAgAs & 8278 & P6$_3$/mmc & $2.9\times10^5$ ($\Gamma$ to A, CB) & $7.6\times10^4$ (A to $\Gamma$, CB ) \\ \hline
        BaMg$_2$Bi$_2$ & 100049 & P$\bar{3}m1$ & $8.9\times10^5$ ($\Gamma$ to A, CB) & $4.5\times10^4$ ($\Gamma$ to A, VB) \\ \hline
        Ca$_2$AsI & 166534 & R$\bar{3}m$ & $1.3\times10^5$ (T to $\Gamma$, CB) & $3.8\times10^4$ ($\Gamma$ to F$_2$, CB) \\ \hline
        Cd$_3$As$_2$$^{\dagger}$ & 107918 & P4$_2$/nmc & $5.8\times10^5$ ($\Gamma$ to Z, CB) & $7.1\times10^4$ (Z to $\Gamma$, CB) \\ \hline
        CoAs$_3$$^{\dagger}$ & 9188 & lm$\bar{3}$ & $4.7\times10^5$ (P to $\Gamma$, VB) & $8.7\times10^4$  ($\Gamma$ to N, CB) \\ \hline
        CrGa$_4$ & 626026 & lm$\bar{3}$m & $2.3\times10^5$ ($\Gamma$ to H$_1$, CB) & $2.6\times10^4$ (P$_1$ to $\Gamma$, CB) \\ \hline
        IrSb$_3$$^{\dagger}$ & 44717 & lm$\bar{3}$ & $4.0\times10^5$ (P$_1$ to $\Gamma$, VB) & $1.1 \times 10^5$ ($\Gamma$ to H$_1$, CB) \\ \hline
        KMgBi$^{\dagger}$ & 616748 & P4/nmm & $6.9 \times 10^5$ (M to $\Gamma$, CB) & $6.5\times10^3$ ($\Gamma$ to Z, CB) \\ \hline
        Na$_3$Bi & 26881 & P6$_3$/mmc & $5.2\times10^5$ ($\Gamma$ to A, CB) & $2.0\times10^4$ (A to $\Gamma$, CB) \\ \hline
        NaCuSe & 12155 & P4/nmm & $5.1\times10^5$ ($\Gamma$ to Z, CB) & $3.7\times10^4$ (Z to $\Gamma$, VB) \\ \hline
        O$_2$Pb & 23292 & P4$_2$/mnm & $8.1\times10^5$ (Z to M, CB) & $8.2\times10^4$ (M to Z, CB) \\ \hline
        RhAs$_3$ & 34052 & lm$\bar{3}$ & $2.7\times10^5$ (P$_1$ to $\Gamma$, VB) & $4.1\times10^4$ ($\Gamma$ to H$_1$, VB) \\ \hline
        RhSb$_3$$^{\dagger}$ & 34049 & lm$\bar{3}$ & $2.7\times10^5$ (P1 to $\Gamma$, VB) & $8.9\times10^4$ ($\Gamma$ to H$_1$, VB) \\ \hline
        Y$_3$GaC & 56396 & Pm$\bar{3}m$ & $7.3\times10^4$ ($\Gamma$ to R, CB) & $1.7\times10^4$  ($\Gamma$ to X, CB) \\ \hline 
        YPdBi$^{\dagger}$ & 616964 & F$\bar{43}m$ & $4.0\times10^5$ (U to $\Gamma$, CB) & $4.2\times10^4$ (U to $\Gamma$, VB) \\ \hline
        $\times$ Cu$_2$Se$^{\dagger}$ & 56025 & Fm$\bar{3}$m & $1.6\times10^5$ ($\Gamma$ to X, CB) & $1.8\times10^4$ (U to $\Gamma$, CB) \\ \hline
        $\times$ GaSb$^\dagger$ & 44328 & F$\bar{43}$m & $1.1\times10^6$ ($\Gamma$ to X, CB) & $5.0\times10^4$  (U to $\Gamma$, VB) \\ \hline
        $\times$ Ge$^{\dagger}$ & 44610 & Fd$\bar{3}$m & 1.6$\times10^6$ ($\Gamma$ to X, CB) & 6.2$\times10^4$ (U to $\Gamma$, VB) \\ \hline
        $\times$ InAs$^\dagger$ & 24518 & F$\bar{43}$m & $8.8\times10^5$ ($\Gamma$ to X, CB) & $3.1\times10^4$ (U to $\Gamma$, VB) \\ \hline
        $\times$ InSb$^\dagger$ & 24519 & F$\bar{43}$m & $8.0\times10^5$ ($\Gamma$ to X, CB) & $4.4\times10^4$ (U to $\Gamma$, VB) \\ \hline
        $\times$ Li$_3$As & 610785 & P$\bar{3}$c1 & $1.0\times10^5$ ($\Gamma$ to A, CB) & $8.4 \times 10^4$ (A to $\Gamma$, CB) \\ \hline
        $\times$ LiCdAs & 609965 & F$\bar{43}$m & $8.6\times10^5$ ($\Gamma$ to X, CB) & $3.0 \times 10^4$ (U to $\Gamma$, VB) \\ \hline
        $\times$ LiMgBi$^{\dagger}$ & 108112 & F$\bar{43}$m & $9.2\times10^5$ ($\Gamma$ to X, CB) & $3.8\times10^4$ (U to $\Gamma$, VB) \\ \hline
        $\times$ Te$^{\dagger}$ & 65692 & P3$_1$21 & $4.1\times10^5$ (H to A, CB) & $3.0\times10^5$ (L to H, VB) \\ \hline

 \end{tabular}
 \caption{Point-node materials identified by our computational search. The chemical formula, Inorganic Crystal Structure Database (ICSD) number, and space group for each compound is listed, as well as the minimal and maximal band velocity at the nodal point, as estimated from DFT data in the TMDB. CB indicates a velocity in the conduction band, and VB indicates the valence band. The $\dagger$ symbol indicates materials for which thermopower and/or magnetothermopower experiments are known by us to have been published. 
 Materials preceded by a $\times$ are those which are known experimentally to not be topological semimetals, but which are listed as topological semimetals in the TMDB.}
 \label{table:searchresults}
\end{table*}

We emphasize that our computational search should not be considered comprehensive due to intrinsic limitations of the TMDB data on which it is based. For example, the database does not included alloyed materials, such as Bi$_{1-x}$Sb$_x$ \cite{pan_magneto-thermoelectric_2025, he_record_2025} or Pb$_{1-x}$Sn$_x$Se \cite{liang_evidence_2013}, which are known to have nearly ideal nodal points and to exhibit strong field enhancement of thermopower. The TaAs family of Weyl semimetals is also prominently missing from Table \ref{table:searchresults}, even though previous DFT-based studies have identified them as Weyl semimetals with Weyl points near the Fermi energy \cite{weng_weyl_2015}. The topological classification of materials in the TMDB is based on certain symmetry indicators, and since the Weyl points in the TaAs family of materials are not protected by any of these symmetries they are (incorrectly) classified as ``trivial'' in the TMDB \cite{vergniory_complete_2019}. The Dirac semimetal ZrTe$_5$ is (probably correctly) classified a bulk insulator, since it has a very small bulk gap \cite{zhang_electronic_2017}, and more recently-synthesized materials like YbMnBi$_2$ (discussed in Sec.~\ref{sec:Sxxexperiments}) are not present at all in the TMDB. We therefore conclude that many, and possibly even the majority, of materials that are promising as topological thermoelectrics will be missed by our search.

It may also happen that materials classified in the TMDB as ``topological semimetals'' are, in fact, band gap insulators. This seems to be the case, for example, for the nine materials listed with a $\times$ symbol in Table \ref{table:searchresults}. All of these materials are actually narrow bandgap insulators, and in fact all of them have previously been identified (either experimentally or computationally) as promising thermoelectric materials (e.g., Refs.~\cite{liu_cu2se_2020,yan_gasb_2024,jaoui_giant_2020, hermans_insb_2001, shchennikov_tellurium_2000, ouyang_LiMgBi_2021, middleton_ge_thermopower_1953, zhong_comprehensive_2019, pallavi_thermoelectric_2023}).  In this sense their appearance in Table \ref{table:searchresults} is evidence of the effectiveness of our computational search, even though the materials are not topological semimetals. Some of these materials are also well-known to exhibit large field enhancement of thermopower. For example, InAs exhibits $S_{xx}$ as high as $10$\,mV/K (or $S_{xx} > 100 k_B/e$) in a 29\,T field \cite{jaoui_giant_2020}, but this enhancement is driven by a field-induced metal-to-insulator transition, in which the chemical potential falls into the band gap, so that the resistivity also rises very strongly with field. The topological bands associated with LiMgBi, Ge, and Te are ``fragile'' (and indicated as such in the TMDB), meaning that they can be easily become gapped by mild perturbations \cite{po_fragile_2018}.

\begin{table*}[!ht]
\centering
 \begin{tabular}{||c c c c c||} 
 \hline
 \textbf{Material} & \textbf{ICSD} & \textbf{Space Group} & \makecell{\textbf{Velocity $\perp$ to} \\ \textbf{Nodal Line} (m/s)} & \makecell{\textbf{Energy Variation} \\ \textbf{Along Nodal Line} (eV) }  \\ [0.2ex] 
 \hline\hline
        KMoS$_3$ & 30752 & P6$_3$/m & 9.0$\times10^5$ & 0.038 \\ \hline
        KMoSe$_3$ & 641256 & P6$_3$/m & 8.3$\times10^5$ & 0.059 \\ \hline
 \end{tabular}
 \caption{Line-node materials identified by our computational search. The chemical formula, ICSD number, and space group for each compound is listed, as well as the velocity perpendicular to the nodal line and the variation in energy of the nodal line. }
 \label{table:nodalsearchresults}
\end{table*}

In addition to these narrow band gap materials, our search does correctly identify a number of materials that are known to be topological thermoelectrics with large thermopower. For example, Cd$_3$As$_2$ and KMgBi have both been shown to exhibit large magnetothermopower (see Table \ref{table:Sxxexp}), and the materials CoAs$_3$, IrSb$_3$, RhSb$_3$, and YPdBi also exhibit reasonably large thermopower (Table \ref{table:Sxxexp}), although to the best of our knowledge there are no published studies of thermopower in a magnetic field for these materials. The skutterudite structure materials CoAs$_3$, RhSb$_3$, RhAs$_3$, and IrSb$_3$ may be particularly promising for future experiments in a magnetic field, since these materials exhibit good nodal points, and CoAs$_3$, RhSb$_3$, and IrSb$_3$ are known to have a Seebeck coefficient ranging from 100 to 500\,$\mu$V/K at room temperature in the absence of field \cite{sharp_thermoelectric_1995_coas3etc}. The materials Na$_3$Bi and KMgBi identified by our search are highly promising in terms of their band structure, and are known to possess good band touching points with both ``fast'' and ``slow'' velocity directions. Future thermopower studies may observe enormous field-enhanced Seebeck effect if the Seebeck coefficient can be measured along the in-plane direction while a magnetic field is applied along the stacking direction. Unfortunately Na$_3$Bi and KMgBi are all highly air sensitive, which limits their practical utility. 

The compound BaMg$_2$Bi$_2$, which appears in Table \ref{table:searchresults}, has been studied by angle-resolved photoemission (ARPES) and confirmed to have a good nodal point at the Fermi energy \cite{takane_bamg2bi2_2021}. Future magnetothermopower studies therefore seem highly promising. Published transport studies of BaAgAs suggest a low carrier density and relatively high mobility \cite{xu_BaAgAs_2020}, but to our knowledge there are no thermopower studies of this material, so that it may also be promising for near-future experiments.

The remaining seven compounds in Table \ref{table:searchresults} are AlY$_3$C, AgAsSr, Ba$_6$Ga$_5$N, Ca$_2$AsI, CrGa$_4$, NaCuSe, and Y$_3$GaC. To our knowledge there are no published ARPES, transport, or thermopower studies of these materials. Thus, given the inherent uncertainties about band structures in the TMDB database, they cannot be recommended as thermoelectrics with a similar level of confidence. Nonetheless, among these candidates the materials NaCuSe and AgAsSr stand out as particularly promising in terms of their band structure. Both exhibit a band touching point at the Fermi level with no other coexisting bands and possess a relatively strong velocity anisotropy that is conducive to magnetic field enhancement.

Our search also uncovers two candidate nodal-line materials, which are listed in Table \ref{table:nodalsearchresults}. These materials both exhibit a relatively long and flat nodal line that does not coincide in energy with other bands. We therefore predict a large field enhancement of the Seebeck effect for these materials, if they can be grown with low intentional doping and if the field is applied in the plane of the nodal line (the $a$-$b$ plane).

We conclude this section with a summary of recommendations for materials indicated by our search to be particularly promising as subjects of future magnetothermopower experiments:

\begin{enumerate}
    \item \textbf{The skutterudite-structure compounds \\ CoAs$_3$, RhSb$_3$, RhAs$_3$, and IrSb$_3$}. Three of these compounds are already known to exhibit relatively large thermopower at room temperature. We predict that the thermopower can be driven substantially higher in the presence of a perpendicular magnetic field.

   \item \textbf{Na$_3$Bi and KMgBi}. These materials are known to be good nodal semimetals with a strong anisotropy of velocity. If a measurement of the Seebeck effect can be performed while a magnetic field is applied along the stacking direction, then they will likely exhibit a strong magneto-Seebeck effect. Such a measurement is challenging because both materials are highly air sensitive. 
   
   \item \textbf{BaMg$_2$Bi$_2$ and BaAgAs}. These materials have been relatively well characterized either by ARPES (BaMg$_2$Bi$_2$) or transport (BaAgAs), and the results are consistent with a nodal band structure that enables a strong magnetothermoelectric effect. To the best of our knowledge there are no published thermopower measurements.

    \item \textbf{NaCuSe and AgAsSr}. These materials, to the best of our knowledge, have never been studied by transport or photoemission. Their band structure looks highly promising for topological thermoelectrics with magnetic field enhancement.

    \item \textbf{KMoS$_3$ and KMoSe$_3$}. These are nodal-line materials that exhibit uncommonly flat and well-isolated nodal lines, according to data in the TMDB. If they can be synthesized with low intentional doping, then they may exhibit large magnetothermopower when a field is applied within the plane of the nodal line (the $a$-$b$ plane).

\end{enumerate}


\section{Summary and Conclusions}
\label{sec:conclusion}

In this review we have attempted to summarize the key theoretical principles that can be exploited to produce efficient topological thermoelectrics, with a particular focus on the topological semimetals (Weyl, Dirac, and nodal-line), considering both their longitudinal (Sec.~\ref{sec:longitudinal}) and transverse (Sec.~\ref{sec:transverse}) responses, both with a magnetic field and without. For each situation we have summarized optimal design criteria that can guide the selection of thermoelectric materials. 

These summaries make clear that, as thermoelectric materials, the topological semimetals are both excitingly new and comfortingly old. On the one hand, the topological semimetals have features that are qualitatively different from the traditional metals and insulators. They contain topologically protected band touching points or lines that enable the electron system to be a conductor with high mobility even while having vanishing density of states. The structure of their Landau levels exhibits an electron-hole degenerate lowest Landau level, which enables large entropy at low carrier density that can be driven ever higher by a magnetic field. And their filled bands need not be electrically inert, so that they can provide Nernst response even without magnetic field. For these reasons the topological semimetals are not, in general, bound by the same longstanding oppressive limitations on thermoelectric efficiency that have prevented a more widespread adoption of thermoelectrics as replacements for conventional heat engines.

But on the other hand, a careful consideration of the topological semimetals as thermoelectrics leads back to many of the same design principles that have guided the field of thermoelectrics for decades. Indeed, the list of desirable properties at the beginning of Sec.~\ref{sec:search} has generally been long-known: low carrier density, high velocity in the transport direction (high mobility), low velocity in a perpendicular direction (high electronic entropy), low thermal conductivity. What is interesting is that in the context of topological semimetals these same criteria can produce qualitatively new effects, such as the non-saturating field enhancement discussed in Sec.~\ref{sec:WeylSxxB} or the anomalous Nernst effect discussed in Sec.~\ref{sec:magneticweyl}. That topological semimetals are already starting to claim record thermoelectric performance (e.g., $zT \approx 2.6$ at $T = 100$\,K in Bi$_{1-x}$Sb$_{x}$ \cite{he_record_2025}) is perhaps an indication that these same principles are beginning to bear new fruits.

Reflecting this confluence of old and new is the set of results returned by our computational search in Sec.~\ref{sec:search}, summarized in Tables \ref{table:searchresults} and \ref{table:nodalsearchresults}. Our unbiased, high-throughput search for topological thermoelectrics uncovers many long-known legacy materials, such as InAs, Cu$_2$Se and elemental Te, while also pointing toward a specific set of promising new materials. Our primary new result is the set of twelve candidate materials listed at the end of Sec.~\ref{sec:search}, which to our knowledge have never been studied experimentally as thermoelectrics, but which look highly promising for field-enhanced thermopower.

We should emphasize that the combination of topology and thermoelectricity still offers a number of frontiers that remain not just to be incorporated into optimal design but to be understood at a fundamental level. Among these is the growing appreciation of \emph{nonlinear} effects driven by quantum geometry \cite{yu_quantum_2025}, such as the nonlinear anomalous Hall effect associated with a dipole of Berry curvature \cite{Sodemann_Fu_2015}. Recent work has shown that the nonlinear anomalous Hall effect has a family of thermoelectric analogs as well \cite{yang_nonlinear_2025}, which could enable  qualitatively new technological capabilities, such as voltage-controlled thermal switching or temperature-controlled electrical switching. Even at the linear response level, there are a variety of effects that are rapidly developing and which have not been covered in this review, including spin Seebeck and magnon drag effects \cite{watzman_utilizing_2025}, goniopolarity (axis-dependent conduction polarity) \cite{he_fermi_2019, uchida_thermoelectrics_2022, adachi_fundamentals_2025}, and topological surface state responses \cite{mccormick_fermi_2018, he_topology_2025}.

\acknowledgments
The authors thank Josh Goldberger, Nicolas Regnault, Tianze Song, and Arch Williams for helpful conversations.
This work is supported by Center for Emergent Materials at The Ohio State University, a National Science Foundation (NSF) MRSEC through NSF award no.\ DMR-2011876.

\appendix 

\section{Linear magnetoresistance by guiding-center drift}
\label{sec:guidingcenter}

In this appendix we briefly recapitulate a generic argument for linear magnetoresistance in a strong magnetic field (first put forward by Murzin \cite{murzin_s_s_conductivity_1984, murzin_quasi-one-dimensional_2000, murzin_electron_2000} and later refined by Song, Refael, and Lee \cite{song_linear_2015}) based on drift of electrons in a random potential. Let us consider a sufficiently strong magnetic field in the $z$ direction that $\omega_c \tau \gg 1$. In this limit the Hall conductivity $\sigma_{xy} = n e /B$, independent of the electron dispersion. The longitudinal conductivity is given by $\sigma_{xx} = e^2 D_{xx} g(\mu)$, where $g(\mu)$ is the density of states. Understanding the resistivity $\rho_{xx} = \sigma_{xx}/(\sigma_{xx}^2 + \sigma_{xy}^2)$ is therefore equivalent to understanding the value of $D_{xx}$.

Suppose that electrons experience a random disorder potential with typical magnitude $V_0$ (in units of volts) and typical correlation length $\xi$. In a sufficiently strong magnetic field that the electron cyclotron radius is much shorter than $\xi$, electron motion in the $x$ and $y$ direction consists primary of drifting by the $\vec{E}\times\vec{B}$ drift along contours of constant potential energy. The typical electric field associated with the disorder potential is $V_0/\xi$, so that the typical $\vec{E}\times\vec{B}$ drift velocity is $v_d = \left| \vec{E}\times\vec{B} \right| /B^2 \sim V_0/(\xi B)$. The diffusion constant of electrons in the $x$ direction is $D_{xx} \sim v_d \xi \sim V_0/B$ \cite{murzin_electron_2000}. Thus we arrive at $\sigma_{xx} \sim e g(\mu) V_0/B$. If we assume $\sigma_{xx} \ll \sigma_{xy}$ [which is equivalent to an assumption of weak disorder potential, $e V_0 \ll n/g(\mu)$], then $\rho_{xx} \simeq \sigma_{xx}/\sigma_{xy}^2$, or
\be 
\rho_{xx} \sim \frac{g(\mu) V_0 B}{n^2}.
\label{eq:rhoxxlinearapp}
\ee 

Notice that if the density of states $g(\mu)$ is a constant (as it is, approximately, outside the EQL), then Eq.~(\ref{eq:rhoxxlinearapp}) implies linear magnetoresistance. In the EQL, on the other hand, the density of states for a Weyl or Dirac semimetal increases linearly with magnetic field as $g(\mu) \sim g_0 e B/(\hbar^2 v_z)$. If $V_0$ is taken to be a constant, then Eq.~(\ref{eq:rhoxxlinearapp}) produces \emph{quadratic} magnetoresistance:
\be 
\rho_{xx} \sim \frac{g_0 V_0 e B^2}{\hbar^2 v_z n^2}
\label{eq:rhoxxB2}
\ee 

On the other hand, if the disorder potential $V_0$ arises from stray impurity charges within the material, then increased density of states improves the ability of the electron system to screen the disorder and $V_0$ declines with field. The conventional theory of disorder screening gives \cite{skinner_coulomb_2014}
\be 
V_0 \sim \left( \frac{ e^2 n_i^2}{(4 \pi \epsilon_0 \varepsilon)^3 g(\mu)} \right)^{1/4},
\label{eq:V0screened}
\ee 
where $n_i$ is the concentration of charged impurities $\varepsilon$ is the dielectric constant, and $1/(4 \pi \epsilon_0)$ is the Coulomb constant. Combining Eqs.~(\ref{eq:rhoxxlinearapp}) and (\ref{eq:V0screened}) gives
\be 
\rho_{xx} \sim \left( \frac{e^2 n_i^2}{(4 \pi \epsilon_0 \varepsilon)^3} \right)^{1/4} \left( \frac{e^2 g_0}{\hbar^2} \right)^{3/4} \frac{B^{7/4}}{n^2 v_z^{3/4}},
\label{eq:rhoxxB74}
\ee 
so that the resistivity increases as $B^{7/4}$. Quadratic and near-quadratic magnetoresistance similar to Eqs.~(\ref{eq:rhoxxB2}) and (\ref{eq:rhoxxB74}) has been observed, for example, in the EQL of ZrTe$_5$ \cite{zhang_observation_2020} and Bi$_{1-x}$Sb$_x$ \cite{he_record_2025}.



\bibliography{designprinciples.bib}

\end{document}